\documentclass[fleqn,usenatbib]{rasti}

\usepackage{newtxtext,newtxmath}

\usepackage[T1]{fontenc}

\DeclareRobustCommand{\VAN}[3]{#2}
\let\VANthebibliography\thebibliography
\def\thebibliography{\DeclareRobustCommand{\VAN}[3]{##3}\VANthebibliography}

\DeclareRobustCommand{\DE}[3]{#2}
\let\DEthebibliography\thebibliography
\def\thebibliography{\DeclareRobustCommand{\DE}[3]{##3}\DEthebibliography}

\usepackage{graphicx}	
\usepackage{amsmath}	
\usepackage[version=4]{mhchem}  
\usepackage{caption}    
\usepackage{subcaption} 

\title[]{Comparing telluric removal methods in their capability to recover injected exoplanet atmosphere signals with high resolution emission spectroscopy}

\author[M. A. F. Keniger et al.]{
Marcelo Aron Fetzner Keniger,$^{1,2}$\thanks{E-mail: Marcelo-Aron.Fetzner-Keniger@warwick.ac.uk}
Matteo Brogi$^{3,4}$
David Armstrong,$^{1,2}$
Siddharth Gandhi$^{1,2}$
\\
$^{1}$Department of Physics, University of Warwick, Gibbet Hill Road, Coventry CV4 7AL, UK \\
$^{2}$Centre for Exoplanets and Habitability, University of Warwick, Gibbet Hill Road, Coventry CV4 7AL, UK \\
$^{3}$Dipartimento di Fisica, Università degli Studi di Torino, via Pietro Giuria 1, I-10125, Torino, Italy \\
$^{4}$INAF-Osservatorio Astrofisico di Torino, Via Osservatorio 20, I-10025 Pino Torinese, Italy \\
}

\date{Accepted XXX. Received YYY; in original form ZZZ}

\pubyear{2026}

\begin{document}
\label{firstpage}
\pagerange{\pageref{firstpage}--\pageref{lastpage}}
\maketitle

\begin{abstract}
    For ground-based high-resolution spectroscopic exoplanet atmosphere studies, removing the contamination from telluric and stellar lines is a crucial step in the analysis process. Despite that, there is no consensus in the literature on the most appropriate way to carry this out. Typically, tellurics are either directly modelled to a percent-level precision, or blindly detrended via Principal Component Analysis (PCA) algorithms, the latter particularly common at infrared wavelengths. Here, we compare three different detrending methods, PCA, Molecfit and our own fitting algorithm \textit{Astroclimes}, measuring their performance in the context of removing telluric and stellar lines to detect exoplanetary atmospheric signals. We specifically look for \ce{H2O} detections, which are particularly affected by residual, time-correlated variability of the telluric spectrum. We use near-infrared CARMENES observations of the day-side of $\tau$ Bootis b to carry out injection and recovery tests. We find that while PCA can sometimes achieve higher SNR, it comes at the expense of stronger signal degradation. All methods struggle more for injected signals with lower orbital velocities, not just affecting the signal's magnitude but its location in velocity space as well. This behaviour is more prominent for PCA than for \textit{Astroclimes} and Molecfit. These results highlight the importance of understanding the effects of different detrending methods on exoplanetary signals, which can lead to biases when characterising real detections. Finally, we report that our attempts to detect a previously claimed water signal from $\tau$ Bootis b all resulted in non-detections.
\end{abstract}




\section{Introduction}\label{sec_introduction}
In the current paradigm of exoplanet studies, alongside detecting new exoplanets, efforts are increasing towards characterising the exoplanets discovered, for example by learning about their atmospheres. Determining the presence and composition of an exoplanet's atmosphere can shed light not only on the planet's interior structure and habitability \citep{rainer2026}, but also on possible formation and evolution scenarios that led the planet to its current state \citep{kempton2024,snellen2025}. 

The field of exoplanet atmosphere characterisation started when sodium was detected in the atmosphere of HD 209458b \citep{charbonneau2002}, which employed data from the Hubble Space Telescope (HST). HST has then delivered two decades of frontier exoplanet science, both in transmission \citep[e.g.][]{sing2016} and emission \citep[e.g.][]{mansfield2022} spectroscopy. Since its launch, JWST has taken up the mantle of trailblazer in the exoplanet atmosphere characterisation field using space-based observations 
and has already detected the presence of molecules in exoplanet atmospheres that could not be detected before, such as \ce{CO2} \citep{jwstteam2023} and \ce{SO2} \citep{tsai2023,powell2024} in WASP-39b. Other molecules (or the lack thereof) have also been detected by JWST observations on the same exoplanet \citep{ahrer2023,alderson2023,rustamkulov2023} as well as in other exoplanets \citep[e.g.][]{coulombe2023,greene2023,lustigyaeger2023,zieba2023,benneke2024,dyrek2024,madurowicz2025}.

Space-based observations, however, are limited by the level of resolving power they can achieve, because it is harder to launch high-resolution instruments into space due to their larger size and weight. At low spectral resolution, the planet's spectral lines are shallower and blended with nearby lines (from both stellar and planetary origin), thus hindering the chances of unequivocally detecting the molecules in exoplanets' atmospheres \citep{birkby2018}. Ground-based observations have been able to keep up with and complement their space-based counterparts in the study of exoplanet atmospheres due to their larger telescope sizes and much higher resolving power. 

At high spectral resolutions ($R > 25000$), one can resolve the features of each molecular species into individual lines whose pattern is unique to each molecule, thus allowing for the disentanglement of the stellar, telluric and exoplanet spectra \citep{birkby2018}. For example, ground-based observations with IGRINS \citep{yuk2010,park2014} could detect CO in the atmosphere of WASP-18b \citep{brogi2023} when JWST observations could not \citep{coulombe2023}, and observations with CARMENES \citep{quirrenbach14} managed to confirm and characterise in more detail the extended helium atmosphere of WASP-107b \citep{allart2019}, which had been detected from HST observations \citep{spake2018}, but due to the telescope's low resolution, the helium feature remained unresolved and poorly characterised \citep{allart2019}.

For ground-based high-resolution spectroscopy (HRS), the cross-correlation technique \citep{snellen2010} has been a powerful tool used to detect chemical species in exoplanet atmospheres using both transmission \citep[e.g.][]{snellen2010,allart2017,brogi2018,giacobbe2021,borsato2025} and day-side spectroscopy \citep[e.g.][]{birkby2013,dekok2013,brogi2014,webb2022}, the latter of which targets the planet's thermal emission directly and can thus be applied to non-transiting exoplanets as well. Not only that, but this technique can also be used to infer the presence of winds on the planet or learn about its orbital inclination and rotation \citep[e.g.][]{snellen2010,brogi2016}. 

High-resolution cross-correlation spectroscopy (HRCCS) combines the signal from all resolved spectral lines originating from the planet's atmosphere, thus increasing the strength of their detection even though these lines are orders of magnitude weaker than the telluric and stellar spectral lines \citep{brogi2014,sanchez-lopez2019}. Nevertheless, this contamination from the Earth's and the host star's atmospheres must be removed from observational spectra (a process often called ``detrending'') so that the planetary signal can be unveiled. Despite being a crucial step in the analysis process, there is no universal approach adopted in the literature to telluric correction, as different wavelength regions (optical versus infrared) and measurements (e.g. molecular versus atomic species, transmission versus emission spectroscopy) are susceptible to different levels of telluric contamination.

Principal Component Analysis (PCA; see \citealt{dekok2013} for its first application and \citealt{giacobbe2021} for a full description of the method, and \citealt{line2021,guilluy2022,pino2022,basilicata2024,guilluy2025} for examples of more recent applications of the method throughout the literature) is a method that uses linear regression to find common modes over time for each pixel in the spectral time series. Because the planetary signal is Doppler-shifted across pixels during the time series, it is not identified as a common mode for any particular pixel, thus only the telluric and stellar signal, which are essentially stationary in wavelength, are removed \citep{birkby2018}. A similar alternative to PCA is the SYSREM algorithm \citep{tamuz2005,mazeh2007}, which was originally designed to detrend systematic effects in light curves, but has been widely used in the context of removing stellar and telluric lines for exoplanet atmosphere studies (introduced by \citealt{birkby2013}; see also \citealt{boldt-christmas2025,palle2025,parker2025,ramkumar2025,yang2025} for recent examples).
PCA is explained in more detail in Section \ref{sec_removal_process}.

Other methods involve removing the telluric lines from a science spectra by dividing it by a template, which can be a telluric standard star \citep[e.g.][]{vidalmadjar1986,vacca2003} or a synthetic telluric transmission spectrum \citep[e.g.][]{lallement93,bailey2007,seifahrt2010,cotton2014}, although the former is an insufficient technique for HRCCS. \cite{ulmermoll19} reviewed three of the most popular synthetic telluric transmission algorithms in the literature, namely \textit{Molecfit} \citep{smette15, kausch2015}, \textit{TelFit} \citep{gullikson2014} and \textit{TAPAS} \citep{bertaux2014}, and compared them with each other as well as with the standard star method. They concluded that synthetic transmission performs better in the wavelength range dominated by water lines, with \textit{Molecfit} being the most complete package between the three, whereas for the wavelength range dominated by oxygen absorption the standard star method has the advantage.

It should be noted, however, that the observations used in \cite{ulmermoll19}, which come from CRIRES \citep{kaeufl2004}, only cover a short wavelength range (two orders that go from $1.166 - 1.188\ \mu$m and $1.247 - 1.272\ \mu$m, respectively). A later article by \cite{sanchez-lopez2019} attempted to use Molecfit to remove telluric lines from CARMENES spectra, which covers a longer wavelength range ($0.95 - 1.7\ \mu$m), and argued that for this longer wavelength coverage the level of residuals left by Molecfit yielded fit uncertainties that were too large to detect the weak \ce{H2O} features of the planetary signal from HD 209458b, which they could detect when using SYSREM instead. 

There have been at least a few other instances where the telluric removal from Molecfit required further correction \citep{casasayas-barris2019,borsato2025,vaulato2025}, but most of the time the affected regions are simply masked out of the analysis \citep[e.g.][]{hoeijmakers2019,hoeijmakers2020,bello-arufe2022,prinoth2022,prinoth2023,costasilva2024,prinoth2025}. Along with masking out wavelength regions with heavily saturated telluric lines, it is also a common practice to remove deep lines from the analysis, as telluric line codes often struggle to model them \citep{kohler2025}. However, what is considered ``deep'' varies inconsistently throughout the literature (e.g. \citealt{deregt2025} consider deep anything below 70\% transmission, while \citealt{gonzalespicos2024} and \citealt{prinoth2024} use 65\% and 50\%, respectively, and others such as \citealt{hoeijmakers2019,hoeijmakers2020,prinoth2022,prinoth2023,costasilva2024} simply state that the deep lines were masked manually). It should be noted that masking deep lines out of the analysis is also common practice when detrending with PCA/SYSREM \citep[e.g.][]{alonso-floriano2019,sanchez-lopez2019,guilluy2022,basilicata2024,lesjak2025,pelaez-torres2026}.

On the other hand, with methods such as PCA and SYSREM, one must be careful as they can be overly aggressive and remove part of the planetary signal on top of the stellar and telluric features if too many components are included \citep{cheverall2023,palle2025b,snellen2025}. Conversely, if not enough components are included, then some residual contamination may still remain. It is therefore a key aspect to determine the optimal number of PCA/SYSREM components, which can differ on a nightly basis and for each spectral order included \citep[e.g.][]{giacobbe2021,basilicata2024,panwar2024,guilluy2025,snellen2025}. Care must also be taken when choosing a metric to optimise the number of components, as \cite{cabot2019} and \cite{cheverall2023} have shown that this has the potential to bias the detection of planetary signals. Another issue with PCA and SYSREM is that they rely on the planetary signal moving across pixels throughout the observations such that it won't be identified as a common mode for any particular pixel and accidentally removed, therefore these methods are less suitable when working with planets that have smaller orbital velocities \citep{maguire2024,morrissey2025,palle2025b,snellen2025}. In such cases, \cite{cheverall2024} suggested that a longer baseline can potentially work around this issue, as sufficient out-of-transit observations could create enough contrast between the planet signal and the telluric and stellar contamination such that it won't be considered a principal component. It is also clear from their preliminary work on the subject that PCA alters the nature of the exoplanet transmission spectrum by partially spreading its signal even to out-of-transit spectra (see e.g. their Fig. 6, also noted in \citealt{dash2024}), which might complicate further analysis

In this paper, we demonstrate the capabilities of a newly developed synthetic telluric transmission algorithm called \textit{Astroclimes} \citep{fetznerkeniger2025}\footnote{\href{https://astroclimes.readthedocs.io/en/latest/}{https://astroclimes.readthedocs.io/en/latest/}} to remove telluric lines from stellar spectra in the context of exoplanet atmosphere studies and compare it to the performances of PCA as formalised in \cite{giacobbe2021} and Molecfit. Similar studies have been carried out in the past, with \cite{langeveld2021} comparing Molecfit to an empirical method that assumes a linear relation between the telluric lines and airmass (employed, for example, in \citealt{snellen2010,brogi2012,brogi2014}), and \cite{maguire2024} comparing the performance of Molecfit and SYSREM. \cite{langeveld2021} looked specifically at Na detections in high-resolution transmission spectroscopy in the optical and found that Molecfit provides a better correction than the empirical method. \cite{maguire2024} also used observations in the optical and tested both methods in two scenarios with different orbital velocities, finding that for higher orbital velocities, both methods perform similarly, with SYSREM typically doing better, while for lower orbital velocities Molecfit is the better choice, precisely because the proximity of the planetary signal to the telluric lines meant that SYSREM removed part of the signal. Here, we use observations in the near-infrared (NIR) instead, which should pose a bigger challenge to the telluric line removal techniques, as the telluric lines are more prevalent in this wavelength range.

The paper is structured as follows: in Section \ref{sec_data} we describe the data used in this work; in Section \ref{sec_methodology}, we explain how each method was constructed and used to remove telluric and stellar lines; in Section \ref{sec_results_and_discussion} we present our results; and finally in Section \ref{sec_conclusions} we share our conclusions. 

\section{Observations}\label{sec_data}
The target chosen for our analysis is the non-transiting hot Jupiter $\tau$ Bootis b, which has been extensively studied in the literature and has been reported to exhibit an atmospheric signal \citep{brogi2012,pelletier2021}. However, the discussion around the presence of water has remained divided, with some sources claiming detections \citep{lockwood2014,webb2022} and others claiming only very low upper limits \citep{pelletier2021,panwar2024}.

To study $\tau$ Bootis b, we used data from the CARMENES spectrograph \citep{quirrenbach14} mounted at the 3.5m telescope at the Calar Alto Observatory. There are a total of 7 nights containing observations of the planet's day-side emission. Five of these nights were employed by \cite{webb2022}: March 26th 2018, May 11th 2018, March 12th 2019, March 15th 2019 and April 11th 2019, henceforth nights 1-5, respectively. The other two nights, April 21st 2018 and April 14th 2019, where not used by them, presumably because of the lower number of observations taken (15 and 69, respectively) and/or the poorer weather conditions. Here, only nights 1 and 4 were used for the injection and recovery analysis described in Section \ref{sec_inj_rec_tests}, and different combinations of all nights were used when attempting to detect any real signal from $\tau$ Bootis b as described in Section \ref{sec_detect_signal}.

\begin{figure}
    \centering
    \includegraphics[width=\linewidth]{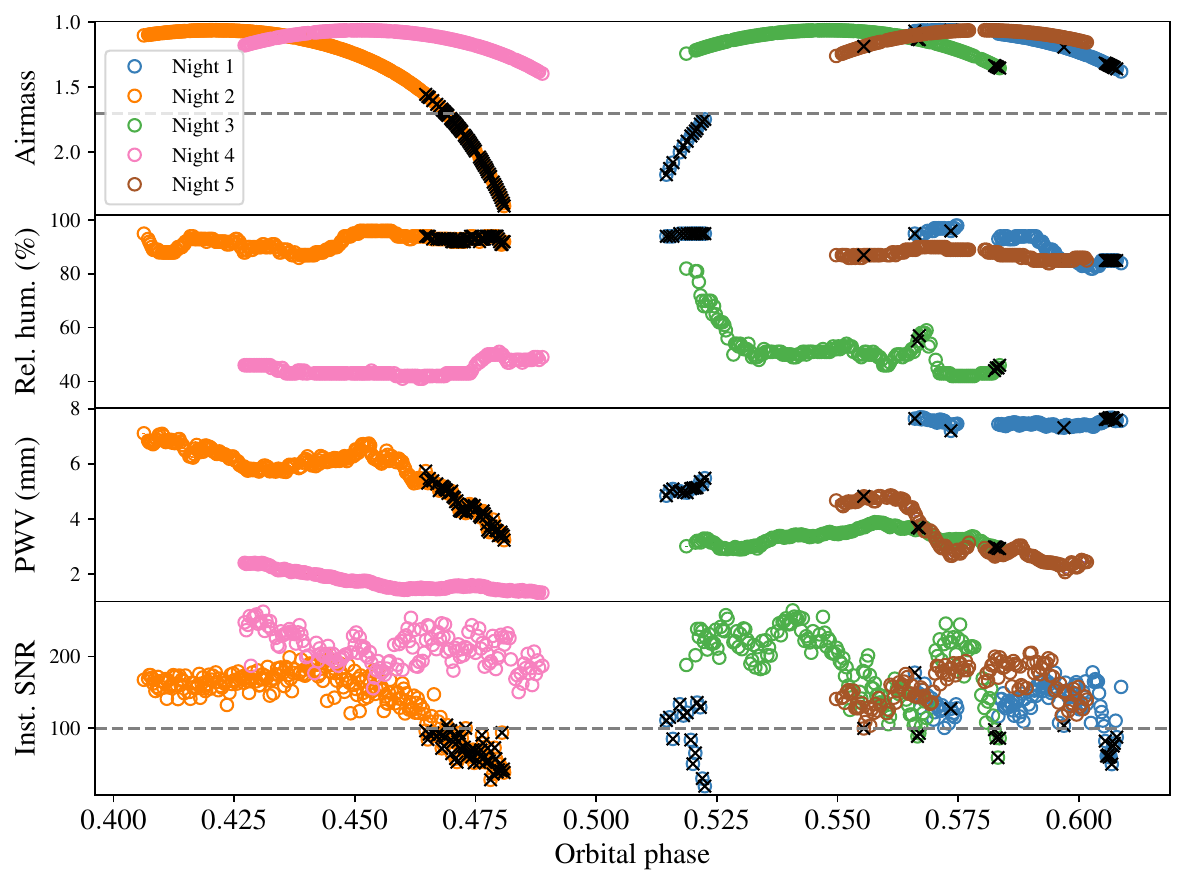}
    \caption{From top to bottom, evolution of the airmass, humidity, precipitable water vapour (PWV) and measured instrumental SNR for the observations of $\tau$ Bootis taken on the nights of March 26th 2018 (night 1), May 11th 2018 (night 2), March 12th 2019 (night 3), March 15th 2019 (night 4) and April 11th 2019 (night 5). The crosses correspond to the points that were removed from the analysis based on the airmass and SNR thresholds, represented here by the horizontal dashed lines on the corresponding panels.}
    \label{fig_obs_weather_conditions}
\end{figure}

Only data from the NIR arm of the spectrograph was used. Out of the 28 NIR orders of CARMENES, orders 9-11 and 18-23 were removed from the analysis for being too densely populated with telluric lines, in line with \cite{webb2022}. Furthermore, the airmass and SNR were used as criteria to filter through the sample. The SNR of each observation was taken as the mean of the measured SNR values for each order, which are stored in the CARMENES FITS header. Observations were limited to those with SNR above 100 and airmass below 1.7. On night 1, there is a clear divide between the spectra within and outside these values, as shown in Figure \ref{fig_obs_weather_conditions}.

Out of the 124 total observations of $\tau$ Bootis taken on night 1, spectra 1-13 and 114-121 plus spectrum 123 were removed due to the airmass and SNR cut, and those can be easily distinguished in Figure \ref{fig_obs_weather_conditions}. However, three other spectra also proved to be outliers, albeit for different reasons. They are spectra 14, 33 and 83. Spectrum 14 exhibited broad absorption features and systematic effects of unknown origin which were present in every order, while spectra 33 and 83 exhibited a variation in their continuum level for orders 15-17 that was not present in any of the other spectra. We speculate that these features are either instrumental in origin or due to issues in the automated data reduction performed before the data is released to the public. These spectra were also removed from the analysis, leaving us a total sample of 99 observations for night 1. Following the same airmass and SNR criteria for the other nights, night 2 had 52 out of 261 spectra removed, night 3 had 6 out of 161, night 4 had zero out of 165, and night 5 had 1 out of 133. No additional spectra had to be removed for other reasons as in night 1.


For the analysis described in the next sections, the observational spectra are organised in ``data cubes'' of dimension ($n_\text{orders}$, $n_\text{observations}$, $n_\text{pixels}$). The number of orders is $n_\text{orders}~=~19$, while the number of observations is $n_\text{observations}~=~99,~209,~155,~165,~132$ (depending on the observing night) and for CARMENES the number of pixels in each spectral order is $n_\text{pixels}~=~4080$. 

\section{Methodology}\label{sec_methodology}
\subsection{Finding the best-fit telluric models}\label{sec_find_best_fit_telluric_model}
The first step to remove telluric lines with \textit{Astroclimes} is to find the best-fit telluric model spectra, as described in \cite{fetznerkeniger2025}. To do that, the abundances of the main molecular constituents in the atmosphere are varied and their best-fit values are obtained from the posterior distributions of a Markov-Chain Monte Carlo (MCMC) that is executed using \texttt{emcee} \citep{emcee}, as described in Section 4 of \cite{fetznerkeniger2025}. A synthetic telluric transmission model is then generated from these best-fit abundance values. Each observational spectrum will have its own best-fit telluric model, thus short-timescale changes in the spectra, such as due to variable water vapour content, can be properly accounted for. 

There are two major differences between the analysis in \cite{fetznerkeniger2025} and the one carried out here. The first one is that in \cite{fetznerkeniger2025}, the only orders used were 12, 16, 25, 26, 27 and 28, but here we use all of the orders not excluded in the previous section. This is because in the previous paper the focus was to use the telluric lines to measure the abundance of greenhouse gases, so the orders chosen are those that contain lines from the main greenhouse gases (\ce{CO2} and \ce{CH4}) and main constituents of the atmosphere (\ce{O2} and \ce{H2O}). For the current work, the priority is to include as much spectral range as possible to optimise the telluric line correction and maximise the planet signal detection.

\begin{table}
    \centering
    \begin{tabular}{|c|c|c|}
        \hline
        Parameter & Value & Reference \\
        \hline
        $T_\text{eff}$ & 6465 K & \cite{soubiran2022} \\
        \hline
        $\log{g}$ & 4.32 cm/s$^2$ & \cite{soubiran2022} \\
        \hline
        [Fe/H] & 0.28 & \cite{soubiran2022} \\
        \hline
        $R_\star$ & 1.42 $R_\odot$ & \cite{borsa2015} \\
        \hline
        $v\sin{i}$ & 16.8 km/s & \cite{luck2017} \\
        \hline
        $\varepsilon$ & 0.328$^{a}$ & \cite{claret1995} \\
        \hline
        $V_\text{sys}$ & -16.4 km/s & \cite{brogi2012} \\
        \hline
        $K_\star$ & 0.4664 km/s & \cite{brogi2012} \\
        \hline
        $P$ & 3.312433 days & \cite{brogi2012} \\
        \hline
        $T_0$ & 2455652.108 BJD & \cite{brogi2012} \\
        \hline
        $K_\text{p}$ & 110 km/s & \cite{brogi2012} \\
        \hline
        $R_\text{p}$ & 1.2 $R_\text{Jup}$ & \cite{webb2022} \\
        \hline
    \end{tabular}
    \caption{Stellar and orbital parameters used in various parts of the analysis process. \\ $^{a}$ Linear limb darkening coefficient in the J band for a star with $T_\text{eff}~=~6500$~K and $\log{g}~=~4.5$~cm/s$^2$}
    \label{tab_stel_orb_pars}
\end{table}

The second difference is that here we are not dealing with a telluric standard star, and therefore the presence of stellar lines cannot be ignored. To deal with that, we incorporate a stellar model in the analysis. The stellar model comes from the Göttingen Spectral Library\footnote{\href{https://phoenix.astro.physik.uni-goettingen.de/}{https://phoenix.astro.physik.uni-goettingen.de/}} and is generated with the stellar atmosphere code PHOENIX \citep{husser2013}. This library provides high-resolution stellar spectra from a discrete grid with different values of effective temperature $T_\text{eff}$, surface gravity $\log{g}$, metallicity [Fe/H] and $\alpha$ element abundance [$\alpha$/M]. These parameters, along with all other stellar and orbital parameters used in this work, are listed in Table \ref{tab_stel_orb_pars}. Based on the selected parameters, the closest match among the Göttingen Spectral Library grid has $T_\text{eff}~=~6500$~K, $\log{g}~=~4.50$~cm/s$^2$, [Fe/H]~=~0.50 and [$\alpha$/M]~=~0\footnote{For models with [Fe/H]~>~0, the only option available in the library for the $\alpha$ element abundance is [$\alpha$/M]~=~0, hence it is not detrimental that this is an unknown quantity for $\tau$ Bootis.}.

To prepare the stellar model spectrum to be incorporated into our analysis, we have to:
\begin{enumerate}
    \item resample its wavelength distribution to match the telluric model spectra and interpolate the flux accordingly via linear interpolation;
    \item modify it to include rotation and instrumental broadening;
    \item shift it to the observational spectra's reference frame by interpolating it to the same wavelength sample;
    \item and normalise it
\end{enumerate}

After resampling the stellar spectrum wavelength distribution to the same one as the telluric model spectra, which has a constant $\frac{\Delta \lambda}{\lambda}$, we interpolate the flux accordingly via linear interpolation. Then, rotation broadening is included by convolving it with a rotation profile given by Equation (18.14) from \cite{gray2005}, reproduced below:

\begin{equation}\label{eq_G}
    G(v) = \frac{2(1 - \varepsilon) \left [ 1 - \left (\frac{v}{v_L} \right )^2 \right ]^\frac{1}{2} + \frac{1}{2} \pi \varepsilon \left [1 - \left (\frac{v}{v_L} \right )^2 \right ]}{\pi v_L \left (1 - \frac{\varepsilon}{3} \right )}
\end{equation}

In the equation above, $\varepsilon$ is the linear limb darkening coefficient, $v_L$ is the maximum velocity shift, given by $v\sin{i}$, and $v$ is the distribution of velocity shifts that we need to compute to calculate the rotation profile, which has to be centred on zero and span at least $\pm v\sin{i}$. Here, we compute it from $\pm(\mathtt{int}(v\sin{i})+5~\text{km/s})$\footnote{\texttt{int} corresponds to the Python integer operator, which for positive numbers rounds down.} to include the wings of the rotation profile, with 1~km/s steps. It should be noted that for $|v|~>~v\sin{i}$, then $G(v)~=~0$. The convolution is calculated using \texttt{NumPy}'s \texttt{convolve} function. 

Instrumental broadening, on the other hand, is included by convolving (again using \texttt{np.convolve}) the spectrum with a Gaussian function whose full-width-half-maximum depends on the resolution of the instrument used. For CARMENES, the resolution on the NIR is reported to be $R~=~80400$ \citep{quirrenbach18}.

Next, to convert the stellar wavelengths from the rest frame $\lambda'$ to the CARMENES reference frame $\lambda$, we use:

\begin{equation}\label{eq_lambda_shift}
    \lambda = \lambda ' \left (1 + \frac{v}{c} \right ) \quad ,
\end{equation}
where $v$ is our target's total radial velocity, in this case the star $\tau$ Bootis, whose radial velocity $RV_\star$ is given by:
\begin{equation}\label{eq_RVstar}
    RV_\star (t) = V_\text{sys}-V_\text{BERV}(t) + K_\star \left \{2\pi\sin{[\phi(t) + 0.5]} \right \}
\end{equation}

In the equation above, $V_\text{sys}$ is the systemic radial velocity of $\tau$ Bootis, $V_\text{BERV}$ is the barycentric Earth radial velocity, in other words, the radial velocity of the Earth with respect to the Solar System's barycentre, and $K_\star$ is $\tau$ Bootis's RV semi-amplitude. $V_\text{sys}$ and $K_\star$ are listed in Table \ref{tab_stel_orb_pars}, whereas $V_\text{BERV}$ is obtained from the CARMENES header for each observation. The systemic velocity is given with respect to the Solar System barycentre, so the $-V_\text{BERV}$ term converts it to the systemic velocity in the Earth's (i.e. the observer's) reference frame. $\phi(t)$ is the orbital phase, given by:

\begin{equation}\label{eq_orb_phase}
    \phi (t) = \frac{t - T_0}{P} \quad ,
\end{equation}

where $P$ is the period and $T_0$ is the time of inferior conjunction. These are measured for the planet $\tau$ Bootis b, so the phase in Equation (\ref{eq_RVstar}) is accompanied by a +0.5 term to account for the $180^\circ$ symmetry between the planet and star orbits.

After being shifted to the observational reference frame, the stellar spectrum is linearly interpolated to the same wavelength sample as the observational spectra. For each of the CARMENES orders, the stellar spectrum is normalised by being divided by its maximum value, such that values are within 0 and 1. The stellar model is then combined with the telluric model via multiplication. \textit{Astroclimes} creates a median filter to normalise the observational spectra, and it uses a dummy telluric spectra combined with the stellar model to exclude the positions containing spectral lines so that only continuum points are included when computing the median filter.

\begin{figure}
    \centering
    \includegraphics[width=\linewidth]{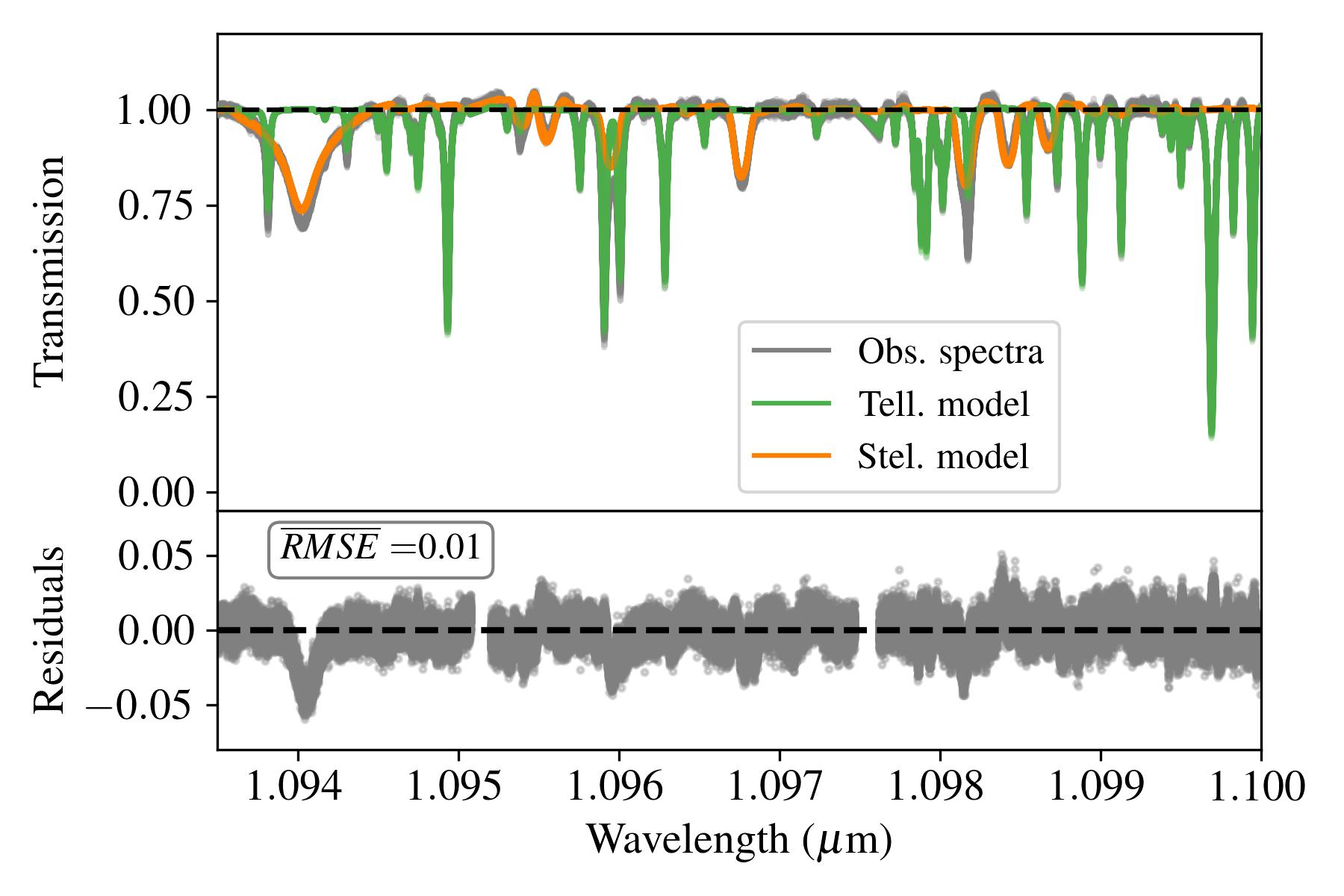}
    \caption{Top: All (good) observations of $\tau$ Bootis from the night of March 26th 2018, with their respective \textit{Astroclimes} best-fit telluric models and the PHOENIX spectrum. Bottom: residuals obtained by subtracting the combined best-fit telluric plus PHOENIX models from the observational spectra.}
    \label{fig_1D_bestfit_models}
\end{figure}

Following the normalisation of the observational spectra, \textit{Astroclimes} then runs an MCMC to find the values for the molecular abundances of \ce{CO2}, \ce{CH4}, \ce{H2O} and \ce{O2} that provide the telluric model that is the best fit for the data. The stellar model spectrum is combined with the telluric model spectra at this stage of the analysis as well, and these combined spectra undergo the same normalisation process as the observational spectra before they are compared to each other. An example of how the best-fit models look compared to their respective data is shown in Figure \ref{fig_1D_bestfit_models} for a short spectral region. We can see that the strongest remaining artifacts are found where the PHOENIX model spectrum is a poor match to the data. However, even though the PHOENIX model is not a perfect match, it provides a better fit than otherwise, so we include it in this MCMC analysis. An additional free parameter could be included in the MCMC to scale the PHOENIX model to provide a better fit, but this possibility has no been explored here. For this reason, we employ a different strategy to remove the stellar lines, as described in Section \ref{sec_removal_process}. The resulting best-fit telluric model spectra (without the PHOENIX model) are also organised in a data cube, with the same dimensions as the observational data cube. 

Molecfit works in a similar way to \textit{Astroclimes}, but there are some key differences. Both algorithms scale an atmospheric profile to find the best-fit column-densities of the relevant molecules, however the atmospheric profiles used and the fitting process are different. While \textit{Astroclimes} uses atmospheric profiles computed with the GGG2020 pipeline \citep{laughner23a} and employs an MCMC to find the best-fit values, Molecfit uses a combination of the GDAS profiles and one of the MIPAS profiles and employs a Levenberg-Marquardt least-squares fit. Because the GDAS profiles contain no information on the molecular abundances as a function of height, and the MIPAS profile used is for a generic latitude and has no temporal information, we believe the GGG2020 profiles provide a more realistic depiction of the state of the atmosphere at the time of observation. Furthermore, an MCMC provides us with more robust results than a least-squares fit, plus the metric we use to quantify the match between model and data (the log likelihood function from \citealt{brogi2019}) is less sensitive to normalisation issues than the $\chi^2$ approach employed by the least-squares fit. Finally, the version of Molecfit used employs the line list database HITRAN 2016, whereas for \textit{Astroclimes} we used HITRAN 2020. For more details about how each algorithm works, we refer the reader to their respective release papers (\citealt{smette15} for Molecfit and \citealt{fetznerkeniger2025} for \textit{Astroclimes}).

It should be noted that Molecfit is developed and typically only works for ESO instruments, for which CARMENES is not. A Molecfit experimental version\footnote{\href{https://support.eso.org/en-GB/kb/articles/molecfit-experimental-version}{https://support.eso.org/en-GB/kb/articles/molecfit-experimental-version}} was developed upon request so that this comparison could be carried out. This experimental version is still incomplete, but works until the Molecfit best-fit transmission spectra are generated, which is sufficient for us. However, because of that, the algorithm crashes at a certain point and requires human input after each spectrum is modelled, which significantly increases the time it takes to compute all the models for one night. For this reason, Molecfit was used only on night 1. The Molecfit models were computed using the default parameters specified by the algorithm. 

\subsection{Computing and injecting a synthetic planetary signal}\label{sec_comp_planet_signal}
Water has been reported in the atmosphere of $\tau$ Bootis \citep{webb2022}. To test the limits for detection with the different detrending methods, we replicate this water signature with a synthetic signal. The planet flux, i.e. the power per unit area and unit wavelength emerging from the planet's surface, is computed for a range of wavelengths. This model contains VMR$_{\ce{H2O}} = 10^{-3.0}$ and is the best-fit employed in \cite{webb2022}, which was computed using the radiative transfer code GENESIS \citep{gandhi17}. 

The planet flux $F_\text{p}$ is combined with the stellar flux $F_\star$ to provide the total flux from the system. The stellar flux is calculated from Planck's blackbody function $B_\lambda (\lambda,T_\text{eff})$ as $F_\star = \pi B_\lambda (\lambda,T_\text{eff})$, and the total flux $F_\text{tot}$ is given by:

\begin{equation}\label{eq_planet_signal}
    F_\text{tot} = 1 + a\left (\frac{F_\text{p}}{F_\star}\right ) \left (\frac{R_\text{p}}{R_\star}\right )^2 \quad ,
\end{equation}
where $R_\text{p}$ and $R_\star$ are the planet and star radii, respectively, and $a$ is a scaling factor that is used to increase the strength of the planetary signal, which we refer to as the ``injected signal strength multiplier'' to avoid confusion with another scaling factor introduced later in the text. From Equation (\ref{eq_planet_signal}), the star flux is 1, and anything extra corresponds to the signal from the planet. 

The planet model is generated in the planet's reference frame, so we need to shift it to the observer's reference frame. To compute the shift using Equation (\ref{eq_lambda_shift}), we calculate the radial velocity of the planet as a function of time, given by the equation below:

\begin{equation}\label{eq_RVplanet}
    RV_\text{p} (t) = V_\text{sys} - V_\text{BERV}(t) + K_\text{p} \sin{[2\pi\phi(t)]} \quad ,
\end{equation}
where $K_\text{p}$ is the planet RV semi-amplitude.

Because the radial velocity of the planet changes over time, as evidenced by the time dependence on the variables $V_\text{BERV}(t)$ and $\phi (t)$, a shifted planetary model is calculated for each observation, such that in the end we will have a planetary model data cube with the same dimensions as the observational data cube. Both cubes are multiplied to inject the planetary signal into the observational spectra. 

\subsection{Removing telluric and stellar lines with \textit{Astroclimes} and PCA}\label{sec_removal_process}
To remove the variations in the light throughput, each spectrum is normalised by its median. Additionally, deep spectral lines (below 0.2 transmission) are masked out alongside bad pixels (i.e. those containing NaNs). In practice, because the data cubes must be kept with fixed dimensions, a boolean ``rule cube'' is created containing the positions of the masked pixels. 

\begin{figure*}
    \centering
    \includegraphics[width=\linewidth]{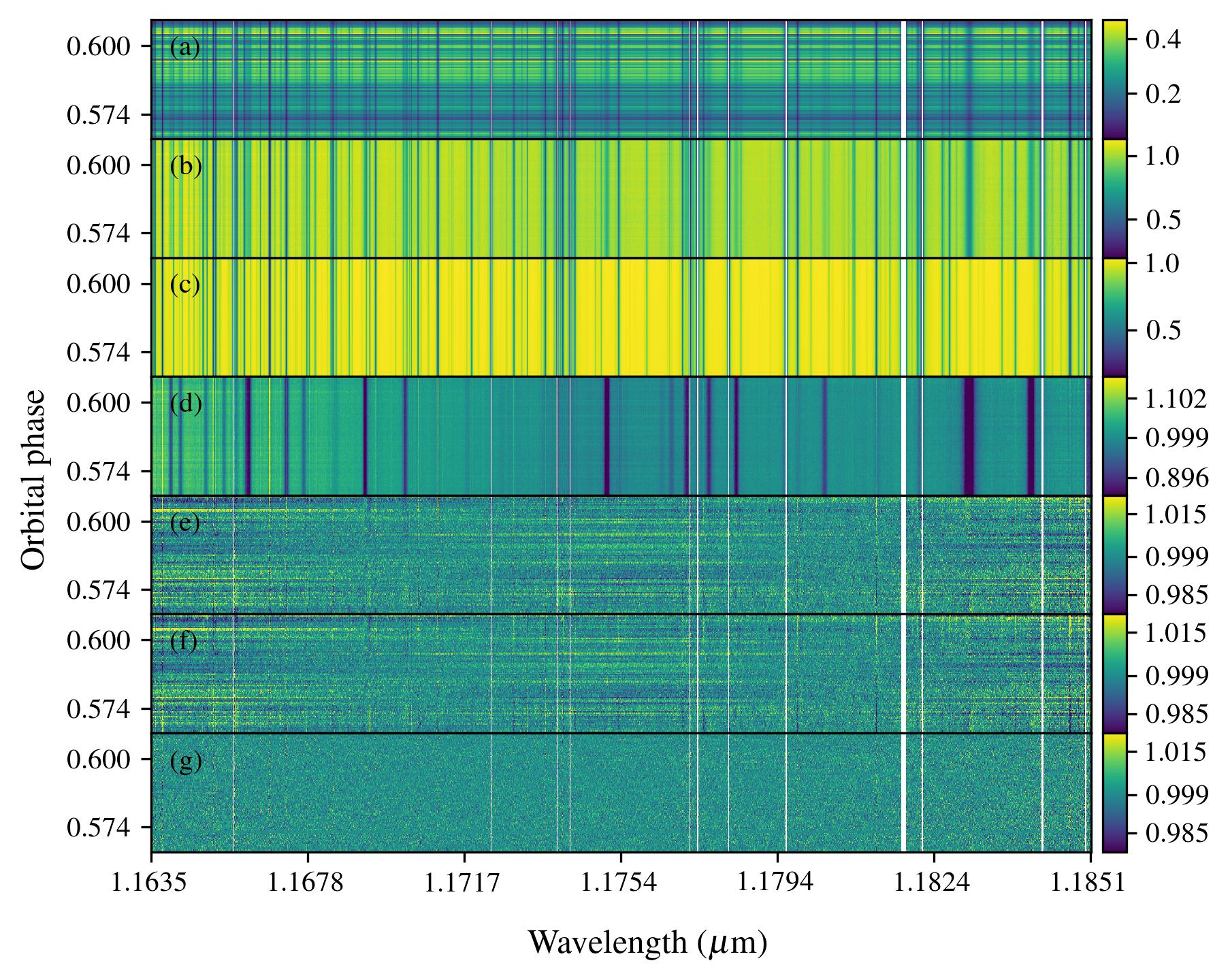}
    \caption{Removal process of telluric and stellar lines for \textit{Astroclimes} for CARMENES NIR order 12. Each row on the $y$-axis corresponds to an individual observation, and the colour bars on the right side correspond to the flux level of each panel. The white vertical lines correspond to the points that were masked out due to being too deep or NaNs. (a) is the observational spectra, (b) is the observational spectra divided its median, (c) is the \textit{Astroclimes} best-fit telluric model, (d) is the result of dividing the median divided observational spectra (b) by the \textit{Astroclimes} telluric model (c) and (e), referred to as the \textit{Astroclimes} residuals, is the result of dividing (d) by the median of (d) over all nights, i.e. the master stellar template, which is done to remove the stellar lines. Panel (f) corresponds to the Molecfit residuals, for which the step-by-step process is exactly the same as for \textit{Astroclimes} and panel (g) corresponds to the residuals when detrending with PCA with $N_C = 3$, for which the step-by-step removal process is illustrated in Figure \ref{fig_PCA_line_removal_process}. For panels (a)-(c), the colour bar limits are the minimum and maximum values of the quantity being plotted, whereas for panels (d) and (e) the limits were taken to be the median $\pm 3\times$ the standard deviation in order to better highlight the noise structure. Panels (f) and (g) have the same colour bar limits as panel (e).}
    \label{fig_astroclimes_line_removal_process}
\end{figure*}

For \textit{Astroclimes}, the next step is to divide the normalised observational spectra by the \textit{Astroclimes} telluric models. This will remove the telluric lines, but the stellar lines will still remain. To remove the stellar lines, we take advantage of the fact that while the planetary signal changes throughout the night, the stellar signal remains roughly the same, so we create a ``master stellar spectrum'' by taking the mean of all (telluric lines removed) observations, and divide each observational spectra by it. The only remaining signal should be the planetary signal and the noise. This process is illustrated in Figure \ref{fig_astroclimes_line_removal_process}. For Molecfit, this process is exactly the same, replacing the \textit{Astroclimes} telluric models by the Molecfit telluric models.

PCA, on the other hand, removes both the telluric and stellar lines in one instance. The application of PCA used here is very similar to that shown in \cite{panwar2024}. This process is done for each order independently, but for all observations simultaneously. It works by calculating the Singular Value Decomposition (SVD) of a standardised version of the normalised observational spectra. This standardised version is obtained by subtracting the mean and dividing by the standard deviation (both done column-by-column, so the median and standard deviation are calculated for each pixel/wavelength value) and ensures that every pixel has the same weight \citep{dash2024} and the PCA is not biased due to the difference in flux between the continuum and the core of deep telluric lines \citep{panwar2024}. From the SVD, the original matrix $A$ (which here corresponds to the observational spectra we are studying) is factorised into three other matrices as:
\begin{equation}\label{eq_SVD}
    A = U\Sigma V^T
\end{equation}

The $U$ matrix contains eigenvectors that are linearly independent and can be used to describe the original matrix $A$, much like a point in Cartesian space can be described by linearly independent vectors pointing at the $x$, $y$ and $z$ directions. In a way, each vector from $U$ describes a certain feature from the original spectra, as illustrated in Figure \ref{fig_PCA_components}. The original matrix can be exactly reconstructed if we use all vectors $U$, given the correct weights. Based on the number of vectors from $U$ we choose to construct our PCA model, we say that the PCA model has that same number of components, which we denote by $N_C$. From the way SVD works, the first component describes the most striking features of the spectra, which in our case would be the telluric and stellar lines, and as we go towards higher components, the features they describe are more and more minute. 

Care must be taken when choosing the number of components to be included in the PCA model, because choosing too few components might not be enough to remove all the unwanted features in the data, but there will come a point when the components start modulating the very signal we wish to detect, so choosing too many components might result in removing part of the planetary signal. Here, we are interested in assessing how an injected signal degrades as we add more PCA components, so we explored several different number of components, as discussed in Section \ref{sec_inj_rec_tests}. Nevertheless, whenever there is need to highlight a specific PCA case for either illustration or comparison purposes, we follow the $\Delta$CCF framework presented in \cite{cheverall2023} to determine the optimal number of PCA components. We note that this approach avoids the biased strategy to optimize PCA on a weakly injected signal, using instead the differential nature of cross correlation to isolate the signal from the noise.

After choosing the number of PCA components, linear regression is used to obtain the weights for the first $N_C$ components of the $U$ matrix that can describe the non-standardised version of the normalised observational spectra. The normalised observational spectra is then divided by the resulting PCA cube, such that again the residuals should only contain the noise and the planetary signal. This process is illustrated in Figure \ref{fig_PCA_line_removal_process}.

\subsection{Using cross-correlation to recover the planet signal}\label{sec_planet_trail_vmap}
As it is common practice in the literature, we used cross-correlation to recover the planetary signal \citep[e.g.][]{birkby2018,snellen2025}. The cross-correlation function (CCF) quantifies the match between the residuals and a planet model which is shifted by a range of radial velocities. Here, we choose this range to be between $\pm 300$~km/s with a step size of 1.5~km/s, totalling $n_\text{RV}~=~401$ different RV values. For each radial velocity we calculate the Pearson's correlation coefficient $C(s)$ (i.e. the CCF) between the shifted model and the residuals:

\begin{equation}\label{eq_PearsoN_Corr_coeff}
    C(s) = \frac{R(s)}{\sqrt{s_{f}^{2}s_{g}^{2}}} \quad ,
\end{equation}
where
\begin{equation}\label{eq_cross_covar}
    R(s) = \frac{1}{N} \sum_{n} f(n)g(n - s) 
\end{equation}
is the cross-covariance between the observed spectrum $f(n)$, which here is the residuals cube, and the template spectrum $g(n-s)$, which here is the shifted planet model cube. $s_{f}^{2}$ and $s_{g}^{2}$ denote the variance of the data and of the model, respectively, given by
\begin{equation}\label{eq_sf2}
    s_{f}^{2} = \frac{1}{N} \sum_{n} f^{2}(n)
\end{equation}

\begin{equation}\label{eq_sg2}
    s_{g}^{2} = \frac{1}{N} \sum_{n} g^{2}(n-s)
\end{equation}

In this context, $N$ is the total number of data points, $n$ refers to each wavelength value in the spectra and $s$ refers to a wavelength shift. 

\begin{figure}
    \centering
    \includegraphics[width=\linewidth]{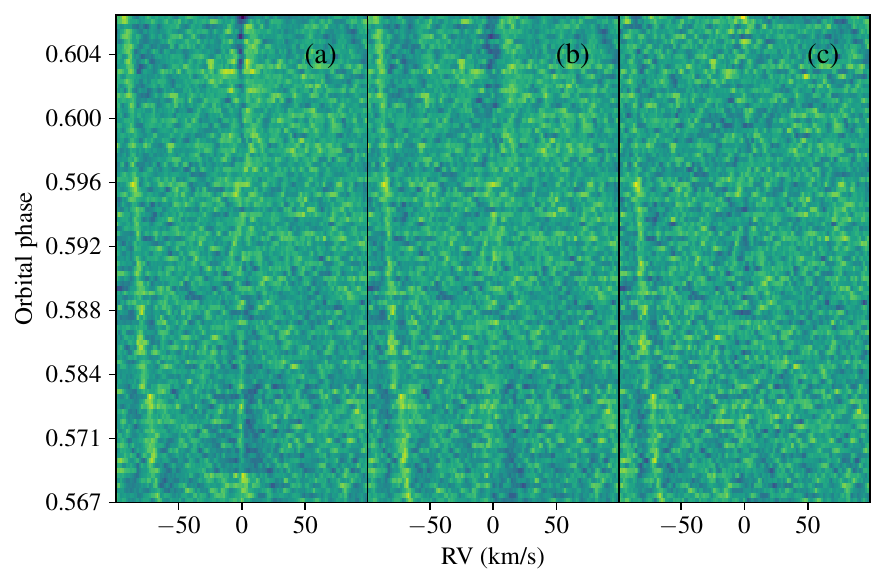}
    \caption{Cross-correlation map in phase-RV space obtained from the residuals when detrending with \textit{Astroclimes} (a), Molecfit (b) and PCA with $N_C = 3$ (c) for an injected planetary signal 10x stronger than the original one. These results are from the injection and recovery test analysis for night 1 with $K_\text{p} = 110$ km/s. The planetary signal corresponds to the slanted line on the left side of each panel. The discontinuity in the planetary signal is due to the gap in observations between orbital phases 0.52 and 0.57, as can be seen in Figure \ref{fig_obs_weather_conditions}. Any vertical structure left at 0~km/s corresponds to leftover telluric signal.}
    \label{fig_planet_trail}
\end{figure}

It is important to note that at this point the rule cube is used to mask out the bad pixels from the calculation of the correlation coefficients. The resulting ``trace matrix'' has the shape ($n_\text{orders}$, $n_\text{observations}$, $n_\text{RVs}$). This matrix is summed over all orders to produce the cross-correlation map in phase-RV space or ``phase-RV map'', examples of which are shown in Figure \ref{fig_planet_trail}, where the planet signal can be clearly seen as the trail between -70~km/s and -90~km/s. It is also common in the literature to refer to this as the ``planet trail''.


\begin{figure*}
    \centering
    \includegraphics[width=\linewidth]{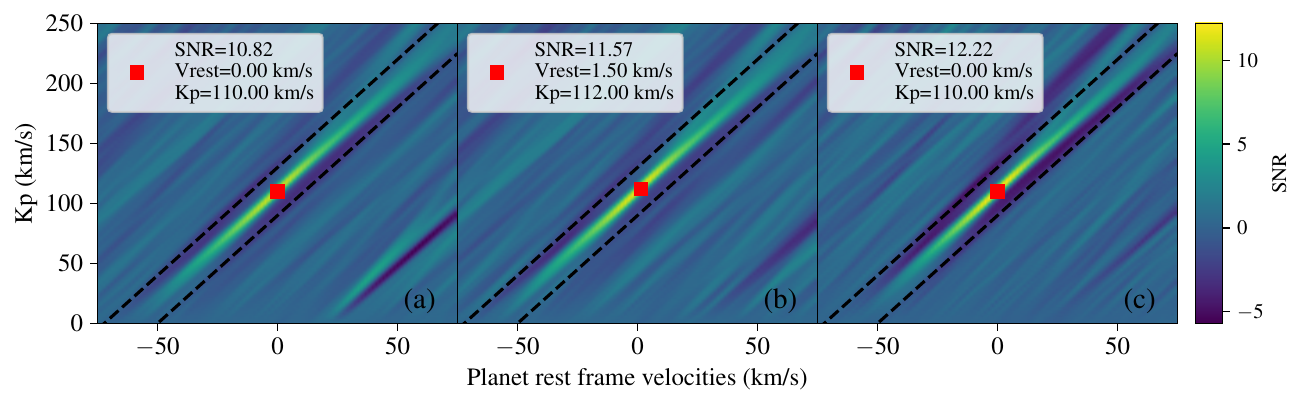}
    \caption{\textit{Left:} Velocity map obtained from the residuals when detrending with \textit{Astroclimes} (a), Molecfit (b) and PCA with $N_C = 3$ for an injected planetary signal that is 10x stronger than the original one. These results are from the injection and recovery test analysis for night 1 with $K_\text{p} = 110$ km/s. The values were converted to SNR as explained in the text. The black dashed lines, defined by visual inspection as $y = 1.8V_\text{rest} + K_\text{p} \pm 20$, delimit the area considered signal, and outside of it is the region considered noise. The red squares correspond to the peak SNR, with the label showing its value and the location it is found.}
    \label{fig_velocity_map}
\end{figure*}

For each of the radial velocities sampled to create the phase-RV map, there are different combinations of $K_\text{p}$ and $V_\text{sys}$ that would result in that radial velocity value. Therefore, we create a grid of $K_\text{p}$ and $V_\text{sys}$ values and evaluate the CCF values in the phase-RV map for each one, which is done via linear interpolation, summed over all observations. The result is what we call the ``velocity map''. Instead of $V_\text{sys}$, we report the results in the planet's rest frame velocity, $V_\text{rest}$, where $V_\text{rest} = 0$~km/s means that the systemic velocity is equal to the literature value used to generate the model. Examples of velocity maps are shown in Figure \ref{fig_velocity_map}, where the values have been converted to SNR by doing $\text{SNR} = \frac{\text{CCF} - \text{noise}}{\text{noise}_\text{STD}}$, where ``noise'' was taken to be the mean of the values outside the area delimited by the two dashed lines (defined by visual inspection), including any leftover artifacts, and ``$\text{noise}_\text{STD}$'' is their standard deviation.


\subsection{Quantifying the effect of the detrending method on the injected signal}\label{sec_effect_on_inj_signal}
After removing the telluric and stellar lines from the observational spectra, ideally we would expect the planetary signal to be intact, but in practice that is not the case, as part of the planetary signal can also be removed or at least altered throughout the detrending process. To quantify how much of the signal was lost, we compare the \textit{Astroclimes}, Molecfit and PCA residuals with a model planetary signal much like the one that was injected in the observational spectra. This model signal has as free parameters the systemic velocity $V_\text{sys}$, the planet semi-amplitude $K_\text{p}$, and a scaling factor $\delta$. The scaling factor is applied in a similar way as the one on Equation (\ref{eq_planet_signal}), but they have different meanings. Before, the signal strength multiplier $a$ was used to strengthen the planetary signal to test different signal strengths, whereas now the scaling factor $\delta$ is being used to determine how much of the original injected signal was lost. Therefore, in this context, we expect $\delta$ to be below one, with values close to one meaning that little of the signal was lost, and values close to zero meaning that almost all the signal was lost. 

To find the best-fit values for the free parameters, we sampled the parameter space by running an MCMC, again using \texttt{emcee}. The MCMC was setup with 10 walkers and a total of 5000 maximum steps, and the posterior distribution is obtained using a burn-in of half the total amount of steps, and the results are calculated as the median of the posterior distribution, with the uncertainties being its standard deviation. We used uniform priors on all free parameters, with the limits being $-50 < V_\text{sys} < 50$, in km/s, $10 < K_\text{p} < 300$, also in km/s, and $-2 < \log{\delta} < 2$, which is unitless. The metric used to calculate the match between model and data comes from \cite{brogi2019}, where the likelihood function $L$ is built from Pearson's correlation coefficient:

\begin{equation}\label{eq_loglike}
    \log{(L)} = - \frac{N}{2} \left [s_{f}^{2} - 2 R(s) + s_{g}^{2} \right ]
\end{equation}

\section{Results and Discussion}\label{sec_results_and_discussion}
\subsection{Injection and recovery tests}\label{sec_inj_rec_tests}
The detrending process was carried out for several different injected planetary signal strengths, namely for values of the signal strength multiplier $a$ (from Equation \ref{eq_planet_signal}) that ranged from 1-10 in unit steps, plus $a~=~3.25,~3.50, 3.75,~4.25,~4.50,~4.75$, totalling 16 values. The finer step size around 3 and 4 was chosen because that is when the transition between non-detection and detection occurs in the velocity map SNR values, which is typically SNR = 4 or 5 for exoplanet atmosphere studies \citep{snellen2025}. Here, we choose SNR = 4. The line removal process with PCA was carried out with six different number of components, $N_C~=~2,~3,~4,~5,~6~\text{and}~10$. 

For each value of the injected signal strength multiplier $a$, an SNR of the recovered signal was calculated at the location where the signal was injected (red squares in Figure \ref{fig_velocity_map}), and an MCMC as described in Section \ref{sec_effect_on_inj_signal} was run to find the best-fit values of the scaling factor $\delta$, the systemic velocity $V_\text{sys}$ and the semi-amplitude $K_\text{p}$. 

For this part of the analysis, two nights were used, March 26th 2018 and March 15th 2019, i.e. nights 1 and 4. These two nights were used to test the detrending methods at varying weather conditions and level of telluric contamination, which we can see is the case for nights 1 and 4 based on their different PWV as shown in Figure \ref{fig_obs_weather_conditions}. For the reasons described in Section \ref{sec_find_best_fit_telluric_model}, Molecfit was only used for night 1. 

In addition to the injection and recovery tests on different nights, which both employed the reported literature value of $K_\text{p}~=~110~$~km/s, for night 1 we also carried out tests on injected signals with lower orbital velocities, namely $K_\text{p}~=~50,~80~$~km/s. Because PCA relies on the fact that the planet's signal shifts pixels throughout the night so that it is not identified as a common mode, it is likely to be more aggressive on signals from planets with lower orbital velocity. Similarly, if the planet's signal does not change as much throughout the night, more of it will be removed when we use the master stellar template to remove the stellar lines, so we expect \textit{Astroclimes} and Molecfit to struggle more in these cases as well. 

\begin{figure*}
    \centering
    \includegraphics[width=\linewidth]{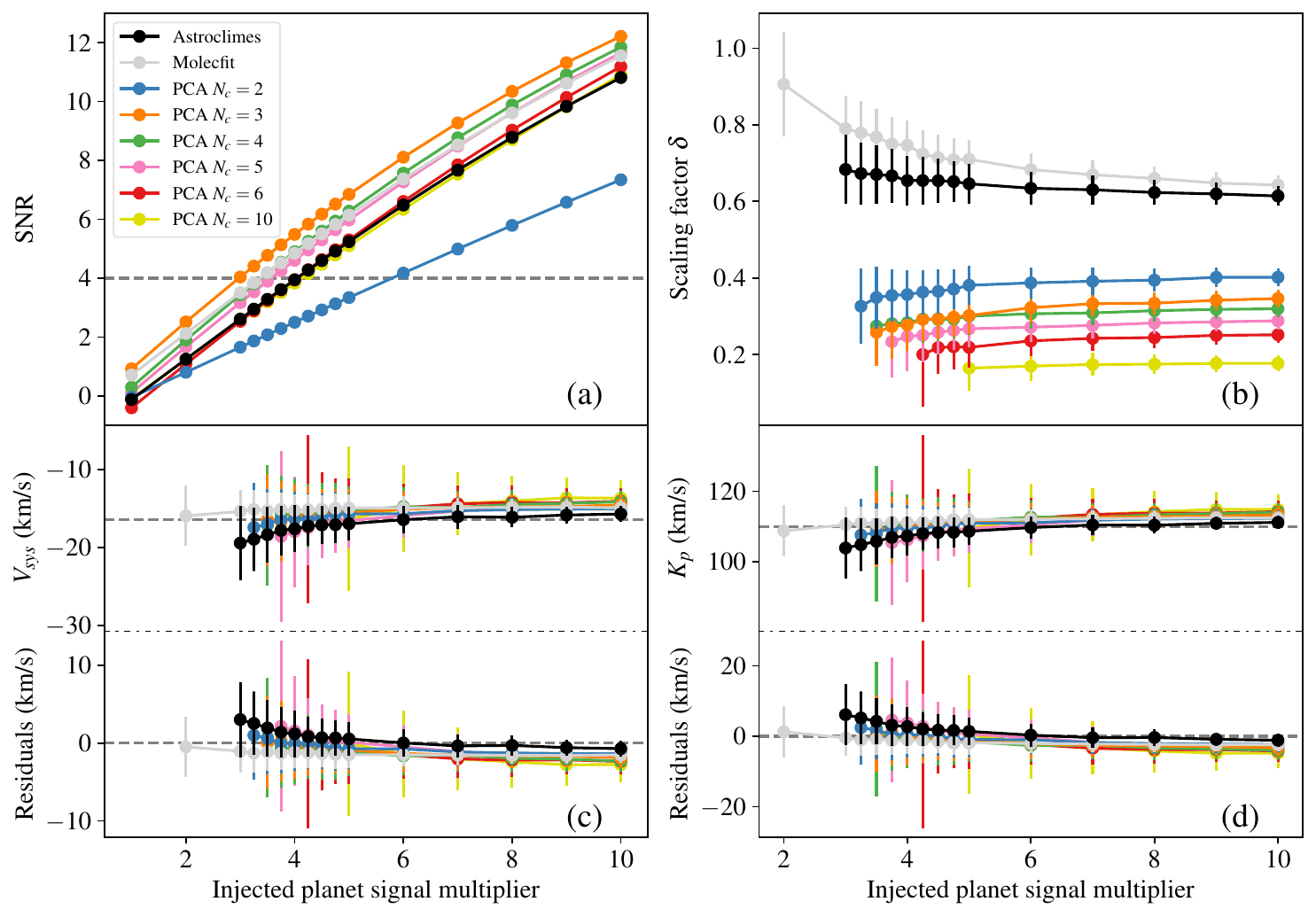}
    \caption{SNR calculated from the velocity map (a), scaling factor $\delta$ (b), $V_\text{sys}$ (c) and $K_\text{p}$ (d) as a function of the injected signal strength multiplier for \textit{Astroclimes}, Molecfit and PCA with 2, 3, 4, 5, 6 and 10 components. These results are for night 1 with $K_\text{p} = 110$ km/s. The horizontal dashed line on panel (a) corresponds to the SNR detection threshold (SNR = 4). For panels (b)-(d), only the results from the MCMC runs that converged are shown, which for \textit{Astroclimes} happened from an injected signal strength of 3x onwards, while for Molecfit it was for 2x and for PCA it was from 3.25x, 3.5x, 3.5x, 3.75x, 4.25x and 5x for $N_C~=~2,~3,~4,~5,~6~\text{and}~10$, respectively. That is why the $x$-axis is sightly different for the left and right panels. The horizontal dashed lines on the top part of panels (c) and (d) correspond to the values of the injected signal, and at the bottom part of these panels their respective residuals are shown, with the horizontal dashed lines marking the zero point. The SNRs reported here are those calculated at the exact position where the signal was injected, which doesn't always coincide with the peak SNR in the velocity map.}
    \label{fig_SNR_MCMC_results_2018_03_26_Kp_110}
\end{figure*}

\begin{figure*}
    \centering
    \includegraphics[width=\linewidth]{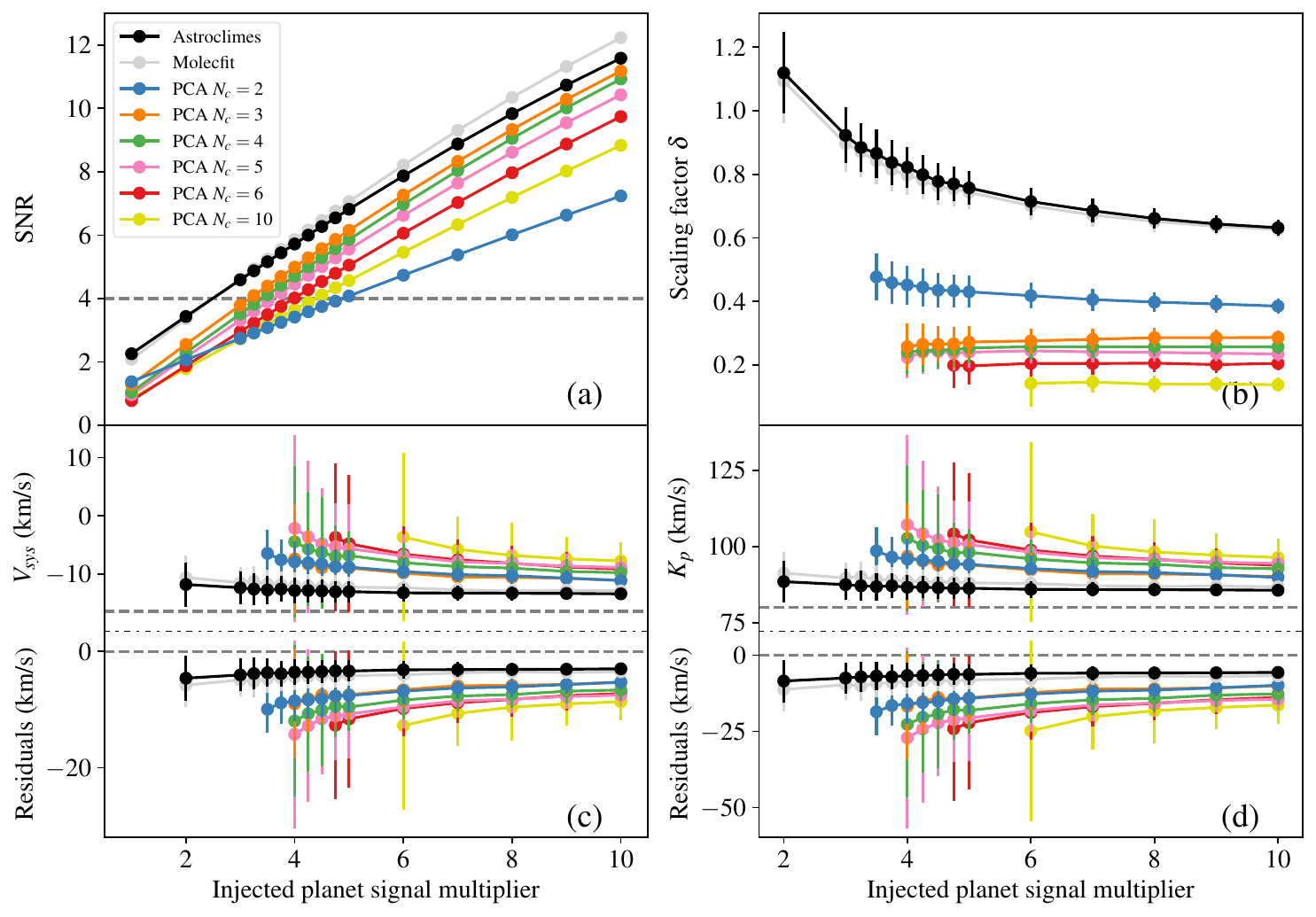}
    \caption{Same as Figure \ref{fig_SNR_MCMC_results_2018_03_26_Kp_110}, but for $K_\text{p} = 80$ km/s. The MCMC runs that converged happened for \textit{Astroclimes} from an injected signal strength of 2x onwards, while for Molecfit it was for 2x and for PCA it was from 3.5x, 4x, 4x, 4x, 4.75x and 6x for $N_C~=~2,~3,~4,~5,~6~\text{and}~10$, respectively.}
    \label{fig_SNR_MCMC_results_2018_03_26_Kp_80}
\end{figure*}

\begin{figure*}
    \centering
    \includegraphics[width=\linewidth]{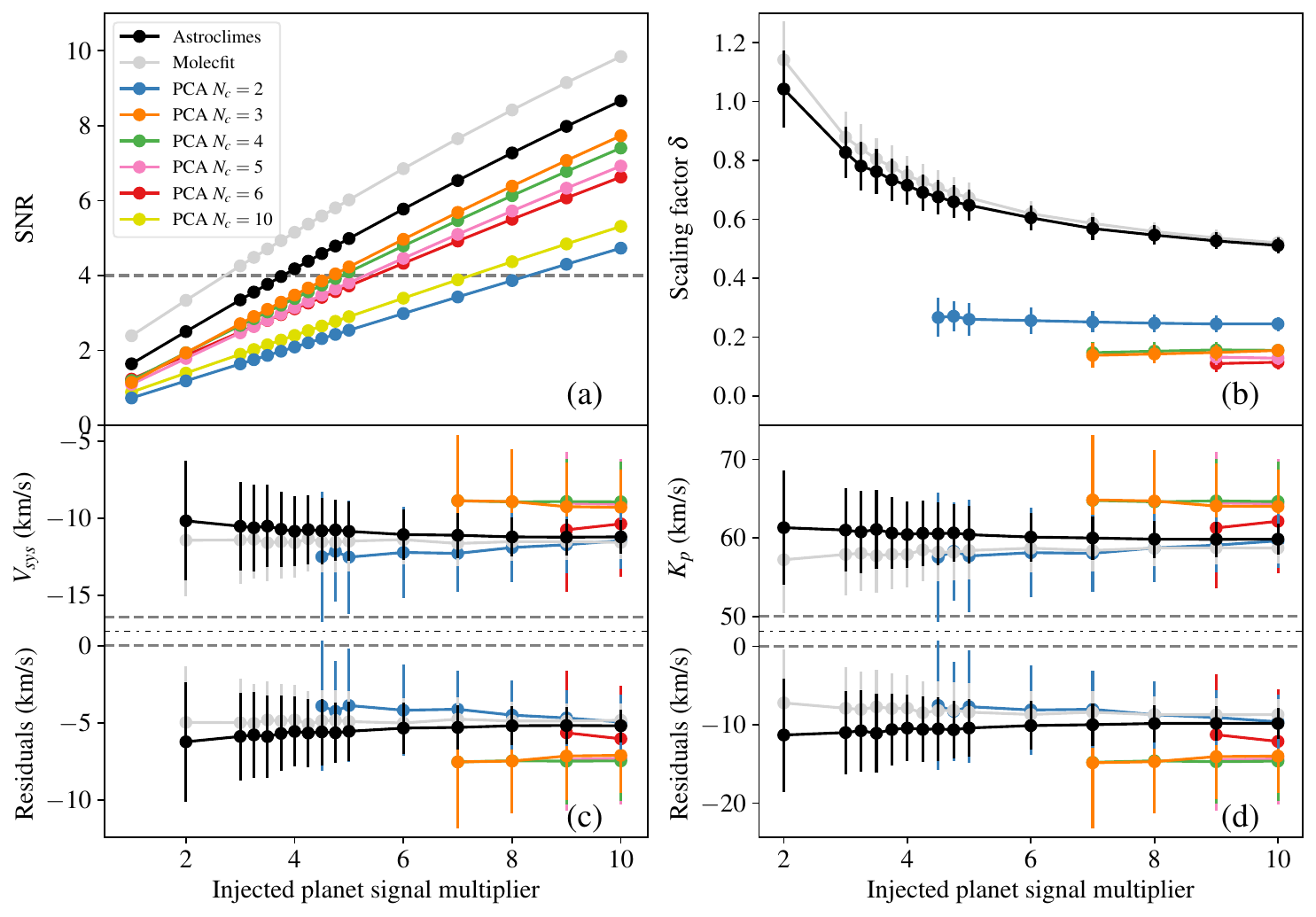}
    \caption{Same as Figure \ref{fig_SNR_MCMC_results_2018_03_26_Kp_110}, but for $K_\text{p} = 50$ km/s. The MCMC runs that converged happened for \textit{Astroclimes} from an injected signal strength of 2x onwards, while for Molecfit it was 2x and for PCA it was from 4.5x, 7x, 7x, 9x, 9x and none for $N_C~=~2,~3,~4,~5,~6~\text{and}~10$, respectively.}
    \label{fig_SNR_MCMC_results_2018_03_26_Kp_50}
\end{figure*}

The results for each of the cases studied for night 1 are shown in Figures \ref{fig_SNR_MCMC_results_2018_03_26_Kp_110}-\ref{fig_SNR_MCMC_results_2018_03_26_Kp_50}, followed by the results for night 2 in Figure \ref{fig_SNR_MCMC_results_2019_03_15_Kp_110}, which are also summarised in Table \ref{tab_inj_rec_results}. In Appendixes \ref{sec_res_night_1_Kp_80}-\ref{sec_res_night_2_Kp_110} we also show their own versions of Figures \ref{fig_planet_trail} and \ref{fig_velocity_map}.

To interpret these results, we need to understand the factors that can affect the injected signal. Much like how the detrending process can alter the depth of the planetary lines, thus changing the magnitude of its signal, it can also distort the shape of the lines, making it seem like the signal is at a different velocity. Hence, besides finding $\delta$ values different than 1, we may find $V_\text{sys}$ and $K_\text{p}$ values that are in contrast with there the signal was originally injected. This behaviour has also been reported by \cite{maguire2024}. Moreover, the noise structure of the velocity map can amount to correlation or anti-correlation with the injected signal, making it seem artificially amplified or dampened, and/or imprinting an apparent velocity shift. Both of these effects are expected to be more prominent when the injected signal is weaker, but they may still be present even for strong injected signals. 

\begin{table*}
    \centering
    \begin{tabular}{|c|c|c|c|c|c|c|c|c|c|c|c|c|}
        \hline
        Night & \multicolumn{9}{|c|}{2018/03/26} & \multicolumn{3}{|c|}{2019/03/15} \\
        \hline
        Injected $K_\text{p}$ & \multicolumn{3}{|c|}{50 km/s} & \multicolumn{3}{|c|}{80 km/s} & \multicolumn{3}{|c|}{110 km/s} & \multicolumn{3}{|c|}{110 km/s} \\
        \hline
        MCMC parameter & $\delta$ & $K_\text{p}$ & $V_\text{sys}$ & $\delta$ & $K_\text{p}$ & $V_\text{sys}$ & $\delta$ & $K_\text{p}$ & $V_\text{sys}$ & $\delta$ & $K_\text{p}$ & $V_\text{sys}$ \\
        \hline
        \textit{Astroclimes} & 61 & 60.22 & -10.97 & 77 & 86.40 & -12.96 & 64 & 109.50 & -16.50 & 101 & 112.23 & -17.43 \\
        Molecfit & 69 & 58.28 & -11.49 & 75 & 88.23 & -12.17 & 70 & 111.73 & -15.00 & N/A & N/A & N/A \\
        PCA $N_C = 2$ & 24 & 59.35 & -11.57 & 40 & 91.78 & -10.10 & 40 & 111.87 & -15.24 & 35$^*$ & 113.17$^*$ & -17.44$^*$ \\
        PCA $N_C = 3$ & 15 & 64.40 & -9.07 & 27 & 92.94 & -9.49 & 31 & 111.26 & -15.53 & 42 & 112.58 & -17.31 \\
        PCA $N_C = 4$ & 15 & 64.68 & -8.90 & 25 & 96.88 & -7.53 & 30 & 111.66 & -15.24 & 37 & 112.88 & -17.45 \\
        PCA $N_C = 5$ & 13 & 64.41 & -9.05 & 24 & 99.48 & -6.19 & 27 & 109.65 & -16.34 & 35 & 113.34 & -17.62 \\
        PCA $N_C = 6$ & 11 & 61.70 & -10.55 & 20 & 98.03 & -6.96 & 23 & 112.21 & -14.97 & 32 & 113.29 & -17.61 \\
        PCA $N_C = 10$ & N/A & N/A & N/A & 14 & 99.35 & -6.26 & 17 & 113.11 & -14.50 & 26 & 113.66 & -17.78 \\
    \end{tabular}
    \caption{Average values of the scaling factor $\delta$, $K_\text{p}$ and $V_\text{sys}$ retrieved by the MCMC runs for each of the cases studied by the injection and recovery tests. The values for $K_\text{p}$ and $V_\text{sys}$ are given in km/s, while $\delta$ is given in percentage. The values reported here correspond to the mean of the MCMC retrieved values for each injected signal strength, but only including the MCMC runs that converged and whose associated SNR is above 4. \\$^*$Based on three converged runs, but none reached SNR above 4.}
    \label{tab_inj_rec_results}
\end{table*}

\begin{figure*}
    \centering
    \includegraphics[width=\linewidth]{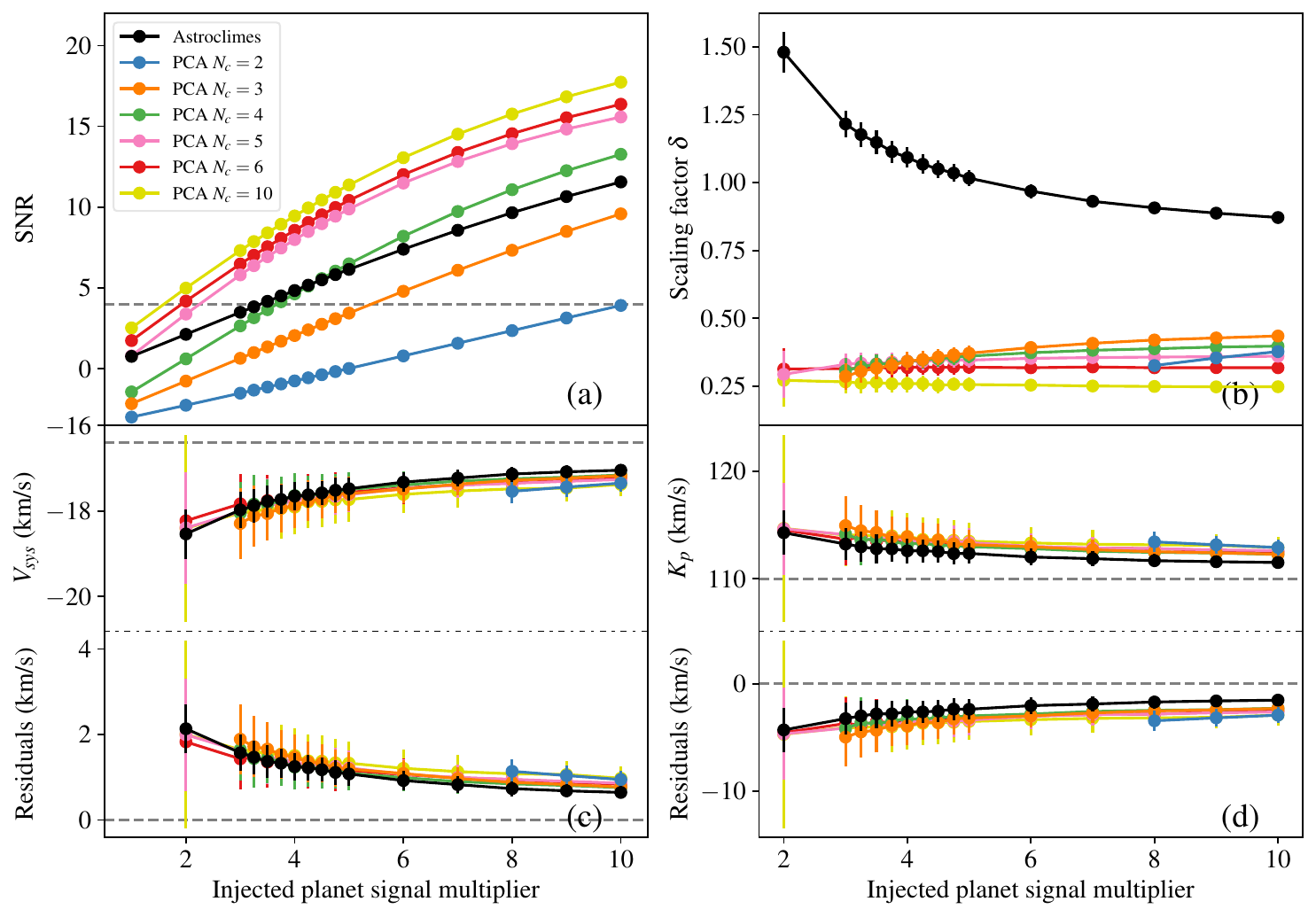}
    \caption{Same as Figure \ref{fig_SNR_MCMC_results_2018_03_26_Kp_110}, but for night 4. The MCMC runs that converged happened for \textit{Astroclimes} from an injected signal strength of 2x onwards, while for PCA it was from 8x, 3x, 3x, 2x, 2x and 2x for $N_C~=~2,~3,~4,~5,~6~\text{and}~10$, respectively.}
    \label{fig_SNR_MCMC_results_2019_03_15_Kp_110}
\end{figure*}

From panel (a) in Figure \ref{fig_SNR_MCMC_results_2018_03_26_Kp_110}, we see that PCA with $N_C~=~3$ achieved the highest SNR, and as we add more components, the calculated SNR goes down. It is worth noting that $N_C~=~3$ is also the optimal number of PCA components for night 1 as determined by the framework from \cite{cheverall2023}. PCA with $N_C~=~2$ is an outlier case because it poorly corrects the telluric and stellar lines, leaving it with a much lower SNR. The SNRs obtained with \textit{Astroclimes} are comparable to those from PCA with $N_C~=~10$, while those obtained with Molecfit are similar to those from PCA with $N_C~=~5$. We speculate that the comparatively lower SNRs from \textit{Astroclimes} are due to residual telluric structure, as evidenced by Figure \ref{fig_velocity_map}, which would increase the noise and thus decrease the SNR. Regardless, the difference in SNR is less than 1.5 and all methods cross the detection threshold at similar injected signal strengths (around 4x for \textit{Astroclimes} versus around 3x for PCA $N_C~=~3$).

Looking at the results for the MCMC scaling factors (Figure \ref{fig_SNR_MCMC_results_2018_03_26_Kp_110}, panel b), significant differences between the methods emerge. The level of signal retention, quantified by the scaling factor $\delta$, is typically much higher for \textit{Astroclimes} and Molecfit than it is for any of the PCA cases. On average, for \textit{Astroclimes}, the detected signal is around 64\% of the original injected signal, whereas for Molecfit it is 70\% and for PCA it is 40\%, 31\%, 30\%, 27\%, 23\% and 17\% for $N_C~=~2,~3,~4~,~5,~6,~\text{and}~10$, respectively, which shows that a forward modelling detrending approach such as \textit{Astroclimes} and Molecfit can preserve more of the planetary signal than PCA.

Panels (c) and (d) from Figure \ref{fig_SNR_MCMC_results_2018_03_26_Kp_110} show that the detected signal location did not deviate from where the signal was injected. This tells us that for this specific case, the detrending methods affect more the depth of the planetary lines compared to their shape. In a real atmospheric retrieval scenario, this would mean that while the signal would be found at the correct location, the retrieved abundance would probably be biased. 

The situation is a bit different when we look at Figures \ref{fig_SNR_MCMC_results_2018_03_26_Kp_80} and \ref{fig_SNR_MCMC_results_2018_03_26_Kp_50}. Now, the SNRs achieved by \textit{Astroclimes} and Molecfit are seemingly higher than any of the PCA cases, with Molecfit taking the top spot this time. Additionally, some scaling factors are above 1, and the $K_\text{p}$ and $V_\text{sys}$ values retrieved by the MCMC do not match where the signal was injected.

The decrease in SNR obtained by PCA is an expected outcome as we move towards lower orbital velocity signals, and while the drop in SNR is mild when we go from $K_\text{p} = 110$ km/s to $K_\text{p} = 80$ km/s, it becomes very apparent when we go to $K_\text{p} = 50$ km/s. Not only is the calculated SNR decreased, but we also see a drastic increase in the signal degradation, evidenced by the much lower $\delta$ values and $V_\text{sys}$ and $K_\text{p}$ values that significantly deviate from the injected values, which highlight just how much the detrending process can bias results. 

For \textit{Astroclimes} and Molecfit, we also see more prominent effects of the detrending method biasing the results for injected signals with lower orbital velocities, particularly when we look at the retrieved $V_\text{sys}$ and $K_\text{p}$ values. The change in $\delta$ is more subtle, and contrary to what we would expect, the signal retention for \textit{Astroclimes} and Molecfit is seemingly higher for $K_\text{p} = 80$ km/s than it was for $K_\text{p} = 110$ km/s. We believe this is at least in part due to the fact that the velocity map coincidentally exhibited a relatively strong positive correlation close to the position where the signal was injected, making it seem amplified. However, the main reason seems to be that for lower injected signals, both \textit{Astroclimes} and Molecfit tend to inflate the level of signal retention, evidenced by the large $\delta$ values that gradually decrease until they converge asymptotically. 

This behaviour is not seen on night 1 for $K_\text{p} = 110$ km/s, but it is very clear for $K_\text{p} = 80,~50$ km/s, and, curiously, also for night 4 (Figure \ref{fig_SNR_MCMC_results_2019_03_15_Kp_110}). For this night, not only were the weather conditions better, but we had a longer baseline of observations with higher instrumental SNR (Figure \ref{fig_obs_weather_conditions}). As a result, \textit{Astroclimes} performed a better telluric correction (see Figures \ref{fig_planet_trail_2019_03_15_Kp_110} and \ref{fig_velocity_map_2019_03_15_Kp_110} versus Figures \ref{fig_planet_trail} and \ref{fig_velocity_map}), so we would expect a better level of signal retention, which is exactly what we see, not just for \textit{Astroclimes}, but overall for PCA as well. However, in addition to the $\delta$ values from \textit{Astroclimes} being inflated due to the reasons described in the previous paragraphs, we believe this night might have two other aspects at play. According to \cite{webb2022}, most of their reported water signal from $\tau$ Bootis b comes from night 4, in which case this could also contribute to the inflation of the retrieved scaling factors. Nevertheless, based on our results from Section \ref{sec_detect_signal}, we cannot claim that with certainty. Finally, if we look at Figure \ref{fig_planet_trail_2019_03_15_Kp_110}, we see that the planet signal crosses the $\text{RV} = 0$ km/s line, so it is possible that residual telluric artifacts may have amplified the signal at that point.

Another interesting aspect of the results from night 4 is the switch in the behaviour of the calculated SNR as function of number of PCA components, where now a larger SNR can be achieved as we add more components. Despite the telluric lines being less prominent for this dataset, it became clear during our analysis that more PCA components were necessary to properly correct the telluric and stellar contamination, which highlights that it is indeed crucial to determine the number of PCA components on a case-by-case basis. Even though $N_C = 10$ achieved the highest SNR, the optimal number of PCA components found by following the framework from \cite{cheverall2023} was $N_C = 5$, which goes to show how differently the detrending parameters can be based on the metric used. From our results, the potential bias imprinted in the retrievals by these two different number of PCA components is only marginal when it comes the location of the signal, but the magnitude of the signal can be further degraded by almost 10\% if we were to choose the higher number of components.

At the beginning of Section \ref{sec_removal_process}, we mentioned that we mask out points below a transmission level of 0.2 from the analysis. We refer to this as the ``deep line threshold''. Throughout this analysis, we also explored the effect that this threshold could have on the results. We found that while using a higher deep line threshold (i.e. removing points at even higher transmission) can sometimes improve the SNR, especially for \textit{Astroclimes}, the MCMC retrieved parameters seem to remain unaffected. This increase in SNR is likely because we remove more points that were not properly corrected by the detrending process, thus decreasing the noise and leading to a higher SNR, and this goes on until there comes a point (i.e. for a deep line threshold over around 0.7) when a significant part of the planetary model is also being removed, so the SNR starts to decrease. However, we stress that these results are not the same for every dataset. For example, the change in the calculated SNR for \textit{Astroclimes} was significant when applying a higher deep line threshold on night 1, when the telluric correction left clear artifacts, but it was not significant for night 4, when the telluric correction was much better. The observed change in the calculated SNR for PCA when trying different thresholds was more subtle and not statistically significant. 

In addition to a simple threshold based on the spectra transmittance, we also explored applying a post-detrending mask that removes highly deviant pixels through a $5\sigma$ clipping, which is a common practice in the literature to mask out pixels that may have not been corrected properly or that suffer from any instrumental artifacts \citep[e.g.][]{brogi2019,boucher2021,webb2022,klein2024,boldt-christmas2025}. Again, we found that while the calculated SNR can increase (less significantly, this time), the retrieved MCMC parameters suffer little to no change. On top of that, we also verified that changing the deep line threshold or applying this post-detrending mask did not change our ability to detect the real signal from $\tau$ Bootis b as explained in Section \ref{sec_detect_signal}. In light of that, we chose to continue with our original setup of a deep line threshold of 0.2, which is a common value found in the literature \citep[e.g.][]{alonso-floriano2019,sanchez-lopez2019,guilluy2022,pelaez-torres2026} and ensures easier reproducibility. We include this discussion here to notify the reader that this is another aspect of the detrending process that has the potential to affect the analysis.

\subsection{Illustrating how the detrending methods affect the planetary signal}\label{sec_res_diff}
\begin{figure*}
    \centering
    \includegraphics[width=\linewidth]{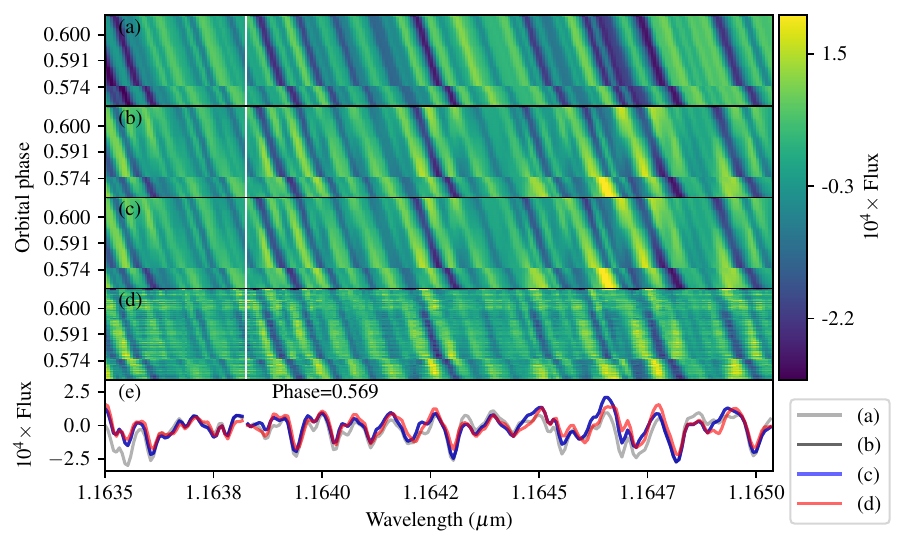}
    \caption{(a) Injected planetary signal which has a strength of 10x the original signal. (b) Median subtracted residuals difference when removing telluric and stellar lines using \textit{Astroclimes} with and without the injected signal. (c) Same as (b), but for Molecfit. (d) Same as (b), but for PCA with $N_C = 3$. Panel (e) shows how panels (a), (b), (c) and (d) look for an individual exposure taken at phase 0.569. The wavelength range covers a fraction of CARMENES NIR order 12. The white vertical lines correspond to the points that were masked out due to being too deep or NaNs. All four panels have the same colour bar limits, which goes from -0.0003 to 0.0002. The results shown are from the injection and recovery analysis for night 1 with $K_\text{p} = 110$ km/s.}
    \label{fig_res_diff}
\end{figure*}

As an additional test to see the effects of \textit{Astroclimes}, Molecfit and PCA on the injected planetary signal, the same detrending process was carried out on the observational spectra without any injected planetary signal. That way, the difference between the residuals from this analysis and the residuals from the analysis with the injected planetary signal in the observational spectra should in theory yield precisely the injected planetary signal, including any alteration due to the applied analysis. This is what is shown in Figure \ref{fig_res_diff}, and we can see that, although not a perfect copy, the residuals difference from \textit{Astroclimes} and Molecfit more closely resemble the injected signal than the residuals difference from PCA, which significantly distorts the injected signal. Reproductions of the same figure with different number of PCA components are shown in Appendix \ref{sec_res_night_1_diff_PCA_comps}, and for the other cases studied, i.e. night 1 with $K_\text{p}~=~80,~50$ km/s and night 2, are shown in Appendixes \ref{sec_res_night_1_Kp_80}, \ref{sec_res_night_1_Kp_50} and \ref{sec_res_night_2_Kp_110}, respectively. Similar effects have been reported by \cite{meech2022}, who compared SYSREM and an airmass detrending approach to a Gaussian-process driven reconstruction of the telluric spectrum. They found that although the former two methods outperform their GP telluric model when it comes to removing stationary contaminating features, they are also a lot more aggressive on the planetary signal, degrading both the wings and the depths of the lines.

Better preserving the planetary signal, albeit at slightly lower SNR, is important because the degradation of the planetary signal can bias retrieved planet parameters \citep{brogi2019}. This degradation is a known caveat of certain detrending methods, and if these methods are to be used, it is important to carry out the same detrending procedure on the models employed in the cross-correlation analysis, so that they experience the same alterations as the data, thus eliminating biases \citep{brogi2019,giacobbe2021,meech2022,maguire2024}. This step is sometimes called ``model reprocessing''. Nevertheless, reprocessing only makes sense when the model spectrum is a close match to the observed (unknown) planet spectrum, because only then the PCA components and distortion effects would be exactly reproduced in both cases. In full retrieval analyses, where the parameter space is sampled in search for the best-fit model, this step is crucial. However, in exploratory cases aimed at detection, such as this work, reprocessing could actually be detrimental. Here, since we inject exactly the model that we recover, reprocessing could lead to an overly optimistic recovery of the planetary signal. Lastly, reprocessing is a computationally expensive process, so for that and the other reasons outlined before, we feel entitled to skip this step in our analyses. 

In the case of \textit{Astroclimes} and Molecfit, most if not all of the alteration of the planetary signal is from the division by the master stellar template. Since it is created by taking the time-average of the (telluric removed) spectra for a given night, such average partially includes the (smoothed) exoplanet spectrum. While they would also require some level of model reprocessing, it would not be as computationally expensive, and we have seen that the level of degradation is typically lower than with PCA. Furthermore, PCA becomes more aggressive on slowly moving planets because their orbital velocities don't vary as much throughout the night, so it would also heavily affect observations in quadrature phases (i.e. phases 0.25 or 0.75). As such, there is certainly benefit in developing alternative strategies that work better in such cases and are not as aggressive on the signal. 

\subsection{Attempt to detect real atmospheric signal}\label{sec_detect_signal}
\begin{figure*}
    \centering
    \includegraphics[width=\linewidth]{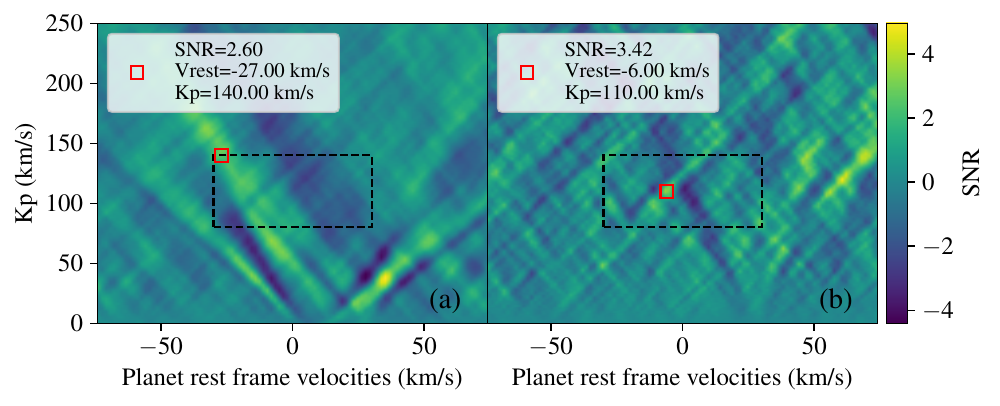}
    \caption{Combined SNR for nights 1-5 from the residuals when detrending with \textit{Astroclimes} (a) and PCA (b). The optimal number of PCA components for each night was 3, 14, 5, 5 and 5, respectively, as determined by the framework from \protect\cite{cheverall2023}. The dashed rectangle delimits the ``signal'' and ``noise'' regions used to calculate the SNR, and its boundaries are $K_\text{p}~=~80,~140$~km/s and $V_\text{rest}~=~-30,~30$~km/s. The open red squares mark the SNR peak.}
    \label{fig_SNR_real_detect}
\end{figure*}

Finally, we wanted to see if our methodology allowed us to independently confirm the water detection reported by \cite{webb2022}. They claim a detection of water in the atmosphere of $\tau$ Bootis b by co-adding all five nights of data and cross-correlating it with a planetary model with VMR$_{\ce{H2O}} = 10^{-3.0}$. We repeated that using both \textit{Astroclimes} and PCA to detrend the data. For each night, the optimal number of PCA components was calculated following the $\Delta$CCF metric from \cite{cheverall2023}, and they are $N_C = 3, 14, 5, 5, 5$ for nights 1-5, respectively. To quantify the strength of a potential signal, again we converted the CCF values to SNR, but this time, instead of determining the ``noise area'' by two slanted lines, we chose to draw a rectangle around the expected signal position and deem everything outside of it to be noise. These results are shown in Figure \ref{fig_SNR_real_detect}.

From Figure \ref{fig_SNR_real_detect}, there is no clear detection of water in the atmosphere of $\tau$ Bootis b. It should be noted that here we are using the orbital solution (i.e. $P$ and $T_0$) for $\tau$ Bootis b from \cite{brogi2012}, whereas in \cite{webb2022} they use the one from \cite{justesen2019}. However, we verified that using that orbital solution also yields a non-detection, and so does not using night 2 in the analysis, which \cite{webb2022} say is the night with the strongest telluric residuals (another hint that this night is problematic is the much higher optimal number of PCA components). Additionally, we searched for a detection by cross-correlating the data with planetary models with VMR$_{\ce{H2O}} = 10^{-4.0}, 10^{-5.0}$, and while it seems that our results favour a solution with VMR$_{\ce{H2O}} = 10^{-4.0}$, the retrieved signal is still not strong enough to warrant a detection. We note that our search for an atmospheric signal was carried out by two similar but independent pipelines, neither of which found a significant detection. 

It is hard to say why we do not find evidence for water in $\tau$ Bootis b when \cite{webb2022} do so using the same dataset. Major differences in our methodologies include the different detrending methods used and the fact that they do model reprocessing when searching for a detection, whereas we do not, for the reasons explained in the previous section.

It is also worth pointing out that the detection of the signal is highly dependent on the proper handling of the orbital parameters for the system. For non-transiting planets such as $\tau$ Bootis b, there can be significant uncertainties in parameters such as the time of inferior conjunction $T_0$, which can lead to uncertainties in other retrieved parameters like $V_\text{sys}$, as was also pointed out by \cite{panwar2024}. In this specific case, using the orbital solution from \cite{brogi2012} versus the one from \cite{justesen2019} results in a difference in phase that corresponds to a few km/s in planet RV. Several different systemic velocities have been reported for $\tau$ Bootis in the literature (e.g. -11.51 km/ from \citealt{webb2022}, -15.4 km/s from \citealt{pelletier2021}, -16.4 km/s from \citealt{brogi2012}, and -17.2 km/s from \citealt{panwar2024}, which used CARMENES, SPIRou, CRIRES and CRIRES data, respectively), with a possible reason for that being an RV shift induced by the wide-orbit binary companion $\tau$ Bootis B \citep{justesen2019,panwar2024}.

\section{Conclusions}\label{sec_conclusions}
The \textit{Astroclimes} algorithm has been recently presented as a new method to measure the abundance of greenhouse gases in the Earth's atmosphere \citep{fetznerkeniger2025}. This is done by modelling the telluric lines in the spectra of telluric standard stars taken with high-resolution spectroscopic observations from ground-based telescopes. Because its aim is to model the telluric lines as best as possible, naturally it can be used to remove the telluric lines as well. This was tested in the context of exoplanet atmosphere studies, for which removal of stellar and telluric lines is an essential step.

Based on observations taken with the CARMENES spectrograph of the $\tau$ Bootis system, which has already shown to exhibit a planetary atmospheric signal \citep{brogi2012,webb2022,panwar2024}, we carried out a series of injection and recovery tests with a synthetic signal containing \ce{H2O} and evaluated how well this signal could be detected when removing the telluric and stellar lines with \textit{Astroclimes}, Molecfit and PCA. 

Two different scenarios of telluric contamination were studied (nights 1 and 4), and for the case with stronger telluric contamination (night 1), three different planet orbital velocities were explored. Due to time constraint reasons, the injection and recovery analysis for Molecfit was only done for night 1.

Overall, we found that PCA tends to be more aggressive on the magnitude of the injected signal compared to \textit{Astroclimes} and Molecfit. The erosion of the signal increases as we add PCA components, which is in line with what was found by \cite{cheverall2023} and \cite{palle2025b}, though due to differences in our methodologies, the level of degradation measured is not directly comparable. The degradation is also more severe when we move towards injected signals with lower orbital velocities, which is an expected outcome based on how PCA works and has also been reported by \cite{maguire2024}. \textit{Astroclimes} and Molecfit, on the other hand, maintained a similar level of signal retention for all three orbital cases studied, though the retention for $K_\text{p} = 80$ km/s was unusually higher than for the other cases, which we attribute to the fact that the signal seemed artificially amplified by the noise structure in the velocity map. Not only that, but it seems that \textit{Astroclimes} and Molecfit tend to inflate the level of signal retention for weak injected signals, a behaviour that is more prominent for injected signals with lower orbital velocities, but can also vary from night to night (see Figure \ref{fig_SNR_MCMC_results_2018_03_26_Kp_110} versus Figure \ref{fig_SNR_MCMC_results_2019_03_15_Kp_110}, panel b). 

The alteration of the injected signal by the detrending process is not limited to the signal's magnitude, but extends to its location as well. While the results for night 1 with a higher orbital velocity signal yielded retrieved $K_\text{p}$ and $V_\text{sys}$ values that matched the injected ones, the bias imprinted by the detrending process became stronger for the lower orbital velocity cases, and that is true for PCA, \textit{Astroclimes} and Molecfit alike. This is understandable as we expect all of these methods to struggle more to remove the telluric and stellar contamination when the planet has a lower orbital velocity. 

Comparing \textit{Astroclimes} with Molecfit, it seems like even though their residuals are very similar, with only minor differences, as can be seen from panels (e) and (f) from Figure \ref{fig_astroclimes_line_removal_process} (and remains true throughout the whole spectral range employed), it seems like \textit{Astroclimes} leaves some strong telluric artifacts on night 1 (Figures \ref{fig_planet_trail} and \ref{fig_velocity_map}). At this point, we could not pinpoint the reason for that. Nevertheless, the results from our injection and recovery analyses showed that \textit{Astroclimes} and Molecfit can achieve similar values of SNR and signal retention, and they seem to bias the injected signal in similar ways. The main advantage of \textit{Astroclimes} is that it can be incorporated alongside the retrieval of the exoplanet's atmosphere parameters. While this might add some computational time, we believe this would make the analysis more robust. At the moment, Molecfit works on ESO's own GUI, which limits flexibility on how it can be applied. Additionally, typically only ESO instruments are supported by their pipelines, so one relies on ESO support to make any changes in the algorithm to suit their specific needs. With \textit{Astroclimes}, we aim to provide an alternative synthetic telluric transmission algorithm which is easy to use and can be adapted to any instrument of choice, both in the optical and infrared. Further work is underway to improve its modelling capabilities and eliminate systematic biases.

Finally, we attempted to detect the water signal on $\tau$ Bootis reported by \cite{webb2022}. We used the same dataset as them, but different detrending methods and no model reprocessing. Despite trying many different configurations, including using the orbital solution from \cite{justesen2019}, which is what was used in \citealt{webb2022}, as well as the orbital solution from \cite{brogi2012}, plus using planetary atmospheric models with varying water abundances (VMR$_{\ce{H2O}} = 10^{-3.0}, 10^{-4.0}, 10^{-5.0}$), we could not detect any significant water signal. This is at odds with what was found in \cite{webb2022}, but it is in line with what was found by \cite{brogi2012} and \cite{panwar2024}. 

Correction of telluric and stellar lines still remains one of the biggest challenges in ground-based spectroscopic observations \citep{cheverall2023,cubillos2025}, especially in the infrared. Our results highlight the importance of understanding the changes that can be imprinted in a planetary signal by the detrending method of choice. If ground-based HRS is to remain in the forefront of exoplanet atmosphere characterisation alongside space-based observations, it is necessary to refine the methods used for cleaning stellar and telluric contamination. 

\section*{Acknowledgements}

We would like to thank the anonymous reviewers for the valuable suggestions and criticism that we believe greatly improved the manuscript. MAFK would also like to thank John Pritchard for implementing support for CARMENES spectra in the MOLECFIT workflow. This research was funded in part by the UKRI (Grants ST/X001121/1, EP/X027562/1) and by the Astronomy and Astrophysics Warwick Prize Scholarship.

\section*{Data Availability}
This article uses or references data from many sources. The molecular cross-sections tables, the atmospheric profiles and the planet emission model were computed by the authors and are not publicly available but can be shared upon reasonable request to the corresponding author. The CARMENES data can be obtained from the CAHA Archive at \href{http://caha.sdc.cab.inta-csic.es/calto/}{http://caha.sdc.cab.inta-csic.es/calto/}. The stellar model comes from the Göttingen Spectral Library, which can be accessed at \href{https://phoenix.astro.physik.uni-goettingen.de/}{https://phoenix.astro.physik.uni-goettingen.de/}.

\section*{Conflict of Interest}
The authors declare no conflict of interest.



\bibliographystyle{mnras}
\bibliography{refs} 




\appendix

\section{Individual PCA components and PCA line removal process}\label{sec_indiv_PCA_comps_PCA_line_rem}
\begin{figure*}
    \centering
    \includegraphics[width=\linewidth]{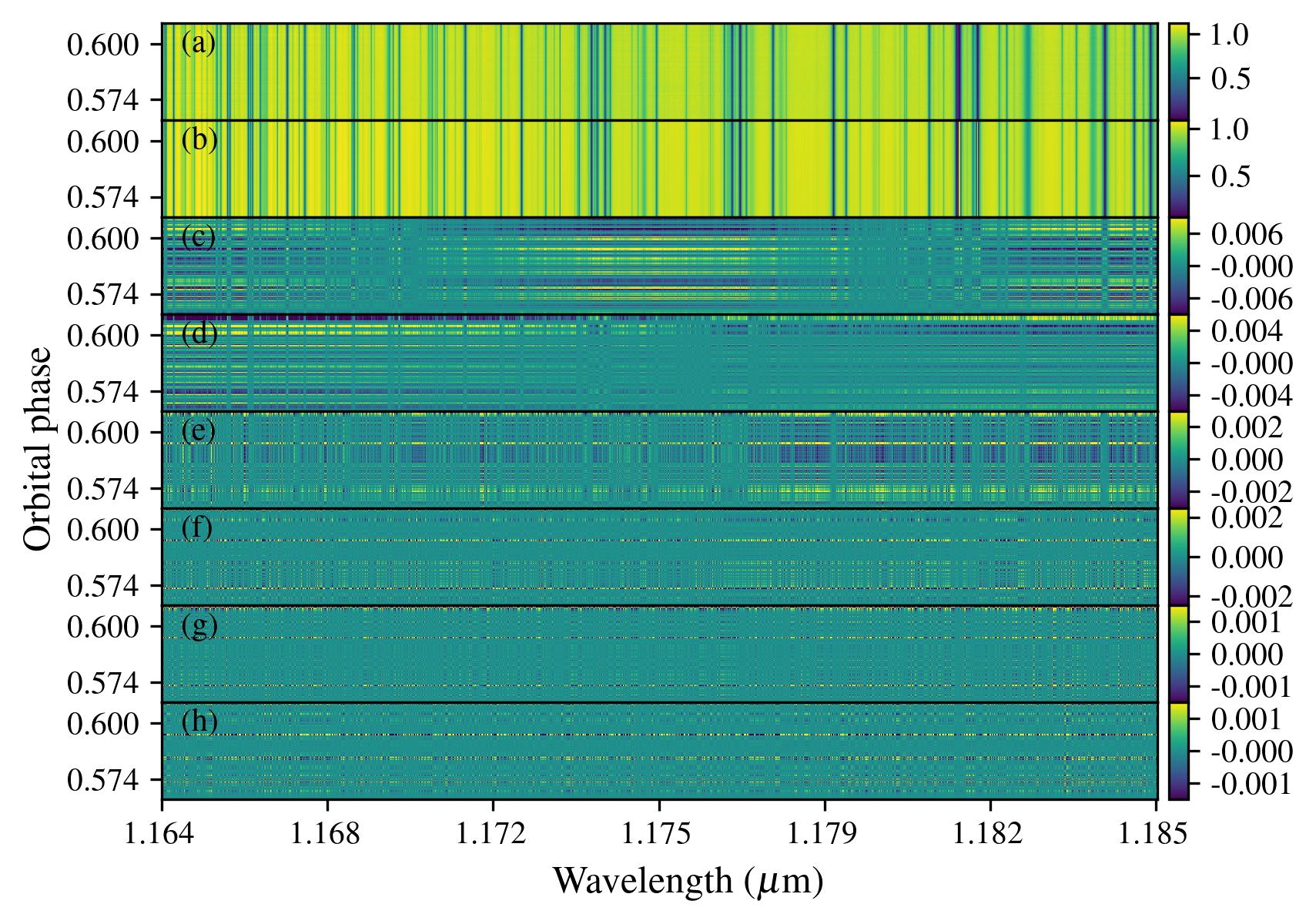}
    \caption{Median divided observational spectra (a) and individual PCA components from 1 to 6 and then 10 (b-h) for the CARMENES NIR order 12 for observations of $\tau$ Bootis taken on the night of March 26th 2018. Each row on the $y$-axis corresponds to an individual observation, represented by the orbital phase, and the colour maps on the right side correspond to the flux level of each panel. For panels (a) and (b), the colour bar limits are the minimum and maximum values of the quantity being plotted, whereas for panels (c)-(h) the limits were taken to be the median $\pm 3\times$ the standard deviation in order to better highlight the structure of each individual component.}
    \label{fig_PCA_components}
\end{figure*}

\begin{figure*}
    \centering
    \includegraphics[width=\linewidth]{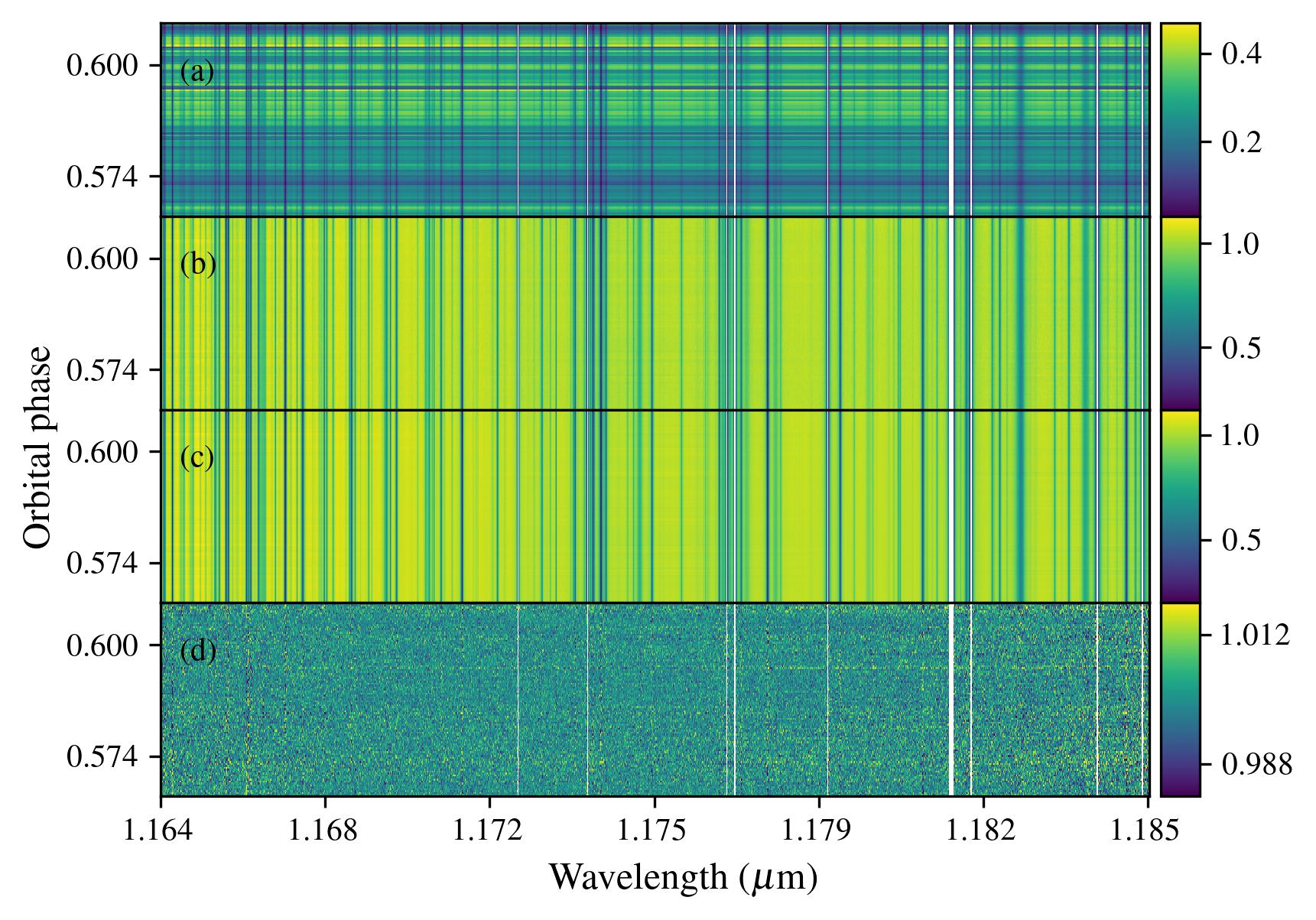}
    \caption{Removal process of telluric and stellar lines for PCA ($N_C = 3$) for CARMENES NIR order 12 for observations of $\tau$ Bootis taken on the night of March 26th 2018. Each row on the $y$-axis corresponds to an individual observation, represented by the orbital phase, and the colour maps on the right side correspond to the flux level of each panel. The white vertical lines correspond to the points that were masked out due to being too deep or NaNs. (a) is the observational spectra, (b) is the observational spectra divided its median, (c) is the PCA model and (d), referred to as the PCA residuals, is the result of dividing the median divided observational spectra (b) by the PCA model (c). For panels (a)-(c), the colour bar limits are the minimum and maximum values of the quantity being plotted, whereas for panel (d) the limits were taken to be the median $\pm 3\times$ the standard deviation in order to better highlight the noise structure.}
    \label{fig_PCA_line_removal_process}
\end{figure*}

Figure \ref{fig_PCA_components} shows the individual contributions of the first 6 PCA components plus for the 10th component, while Figure \ref{fig_PCA_line_removal_process} illustrates the steps involved in removing stellar and telluric lines from spectra using PCA for a case with $N_C = 3$.

\section{Results from injection and recovery tests for night 1 with \texorpdfstring{$\mathbf{K}_\text{p}~=~\mathbf{110}~$km/s}{Kp=110km/s} for different number of PCA components}\label{sec_res_night_1_diff_PCA_comps}
\begin{figure*}
    \centering
    \includegraphics[width=\linewidth]{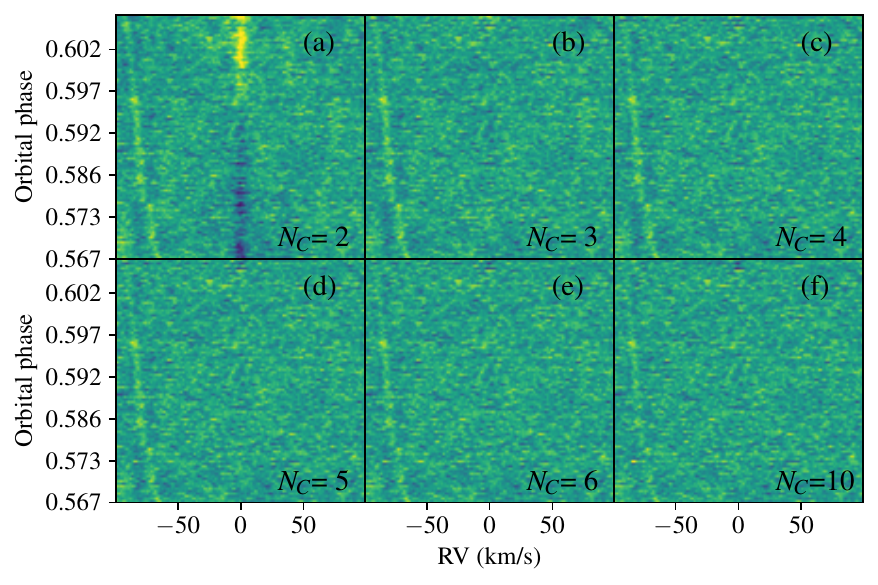}
    \caption{Same as Figure \ref{fig_planet_trail}, but from the residuals when detrending with PCA with $N_C~=~2~\text{(a)},~3~\text{(b)},~4~\text{(c)},~5~\text{(d)},~6~\text{(e)}~\text{and}~10~\text{(f)}$. The data here was injected with a planetary signal that is 10x stronger than the original signal.}
    \label{fig_planet_trail_PCA_comps}
\end{figure*}

\begin{figure*}
    \centering
    \includegraphics[width=\linewidth]{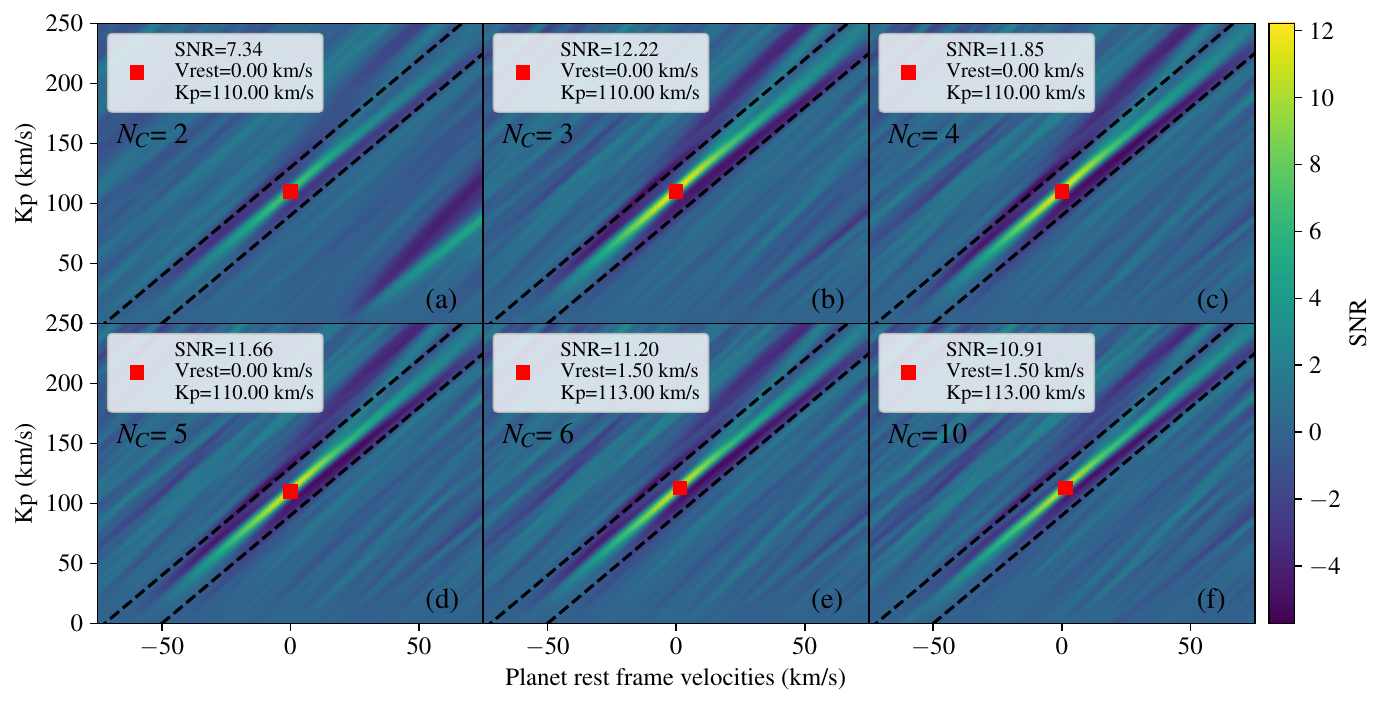}
    \caption{Same as Figure \ref{fig_velocity_map}, but from the residuals when detrending with PCA with $N_C~=~2~\text{(a)},~3~\text{(b)},~4~\text{(c)},~5~\text{(d)},~6~\text{(e)}~\text{and}~10~\text{(f)}$. The data here was injected with a planetary signal that is 10x stronger than the original signal.}
    \label{fig_SNR_maps_PCA}
\end{figure*}

\begin{figure*}
    \centering
    \includegraphics[width=\linewidth]{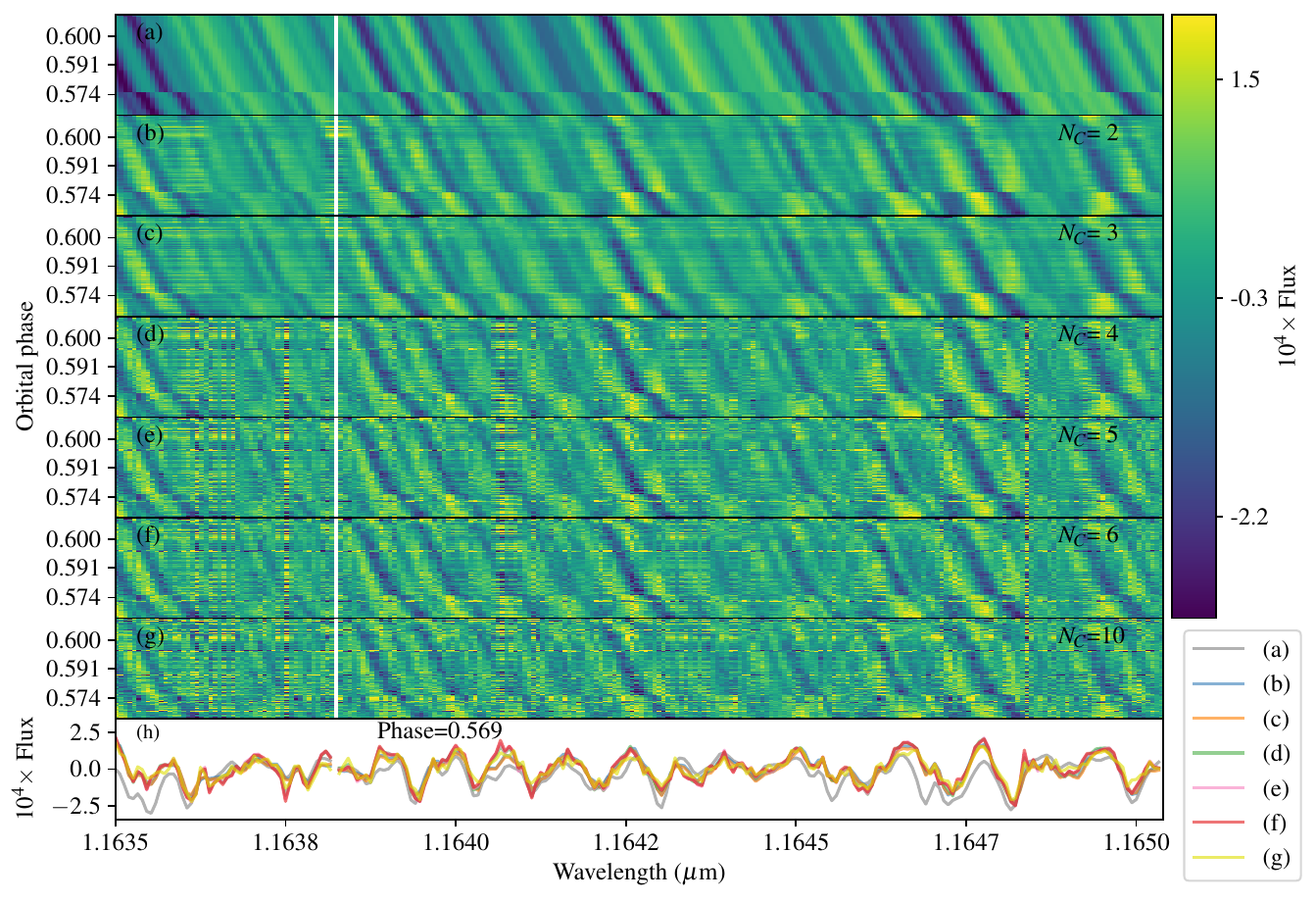}
    \caption{Same as Figure \ref{fig_res_diff}, but now panels (b)-(g) show the residuals difference when detrending with PCA with $N_C~=~2,~3,~4,~5,~6,~10$, respectively.}
    \label{fig_res_diff_PCA_comps}
\end{figure*}

Figures \ref{fig_planet_trail_PCA_comps}, \ref{fig_SNR_maps_PCA} and \ref{fig_res_diff_PCA_comps} show the cross-correlation maps in phase-RV space, the velocity maps and the residuals difference  for PCA with different number of components.

\section{Results from injection and recovery tests for night 1 with \texorpdfstring{$\mathbf{K}_\text{p}~=~\mathbf{80}~$km/s}{Kp=80km/s}}\label{sec_res_night_1_Kp_80}

\begin{figure}
    \centering
    \includegraphics[width=\linewidth]{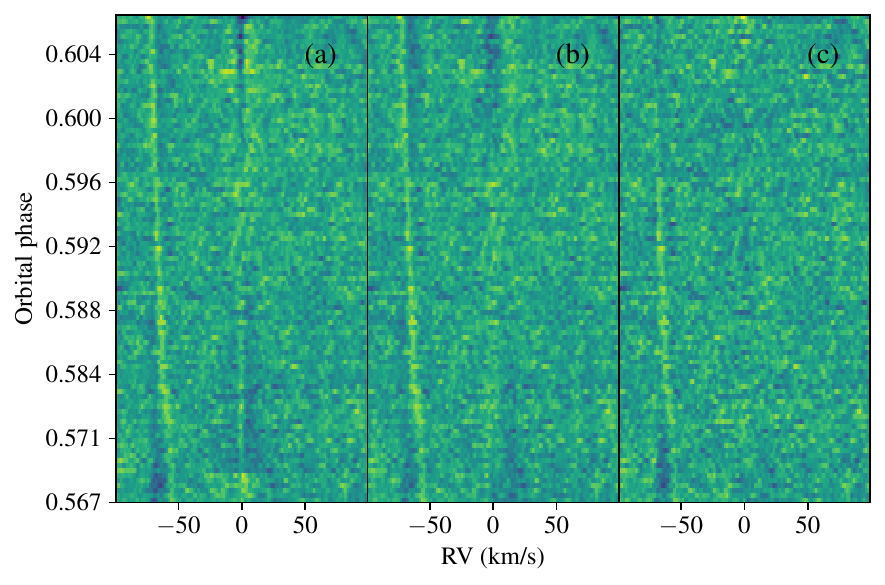}
    \caption{Same as Figure \ref{fig_planet_trail}, but for the results from night 1 with $K_\text{p} = 80$ km/s.}
    \label{fig_planet_trail_2018_03_26_Kp_80}
\end{figure}

\begin{figure*}
    \centering
    \includegraphics[width=\linewidth]{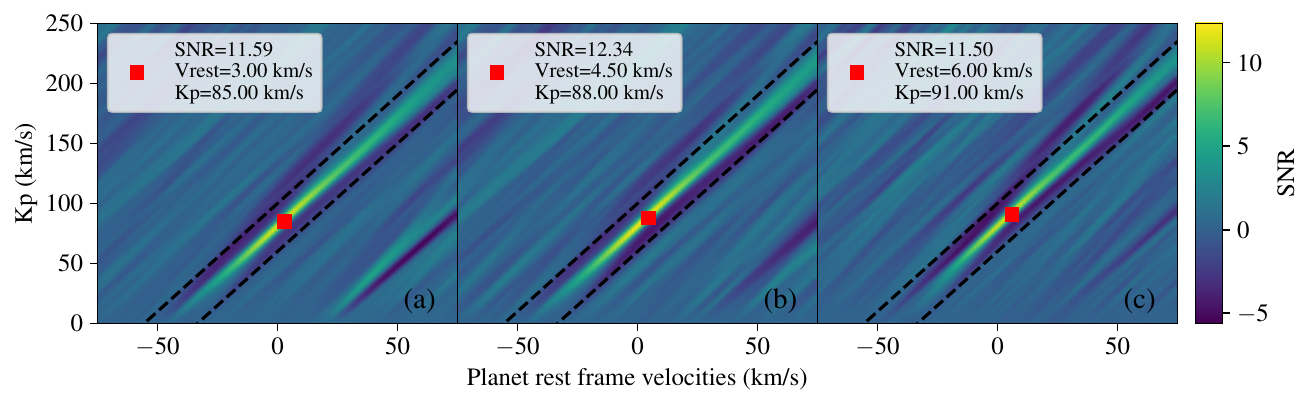}
    \caption{Same as Figure \ref{fig_velocity_map}, but for the results from night 1 with $K_\text{p} = 80$ km/s.}
    \label{fig_velocity_map_2018_03_26_Kp_80}
\end{figure*}

\begin{figure*}
    \centering
    \includegraphics[width=\linewidth]{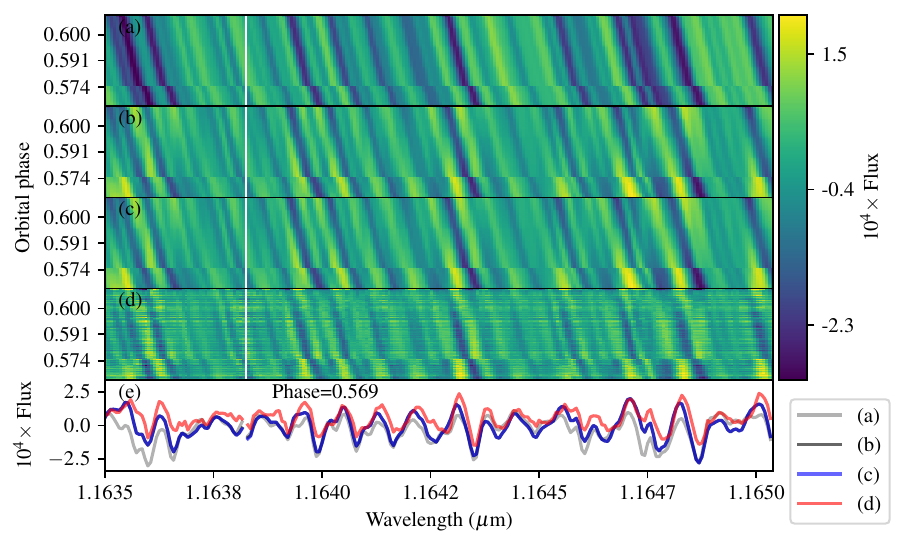}
    \caption{Same as Figure \ref{fig_res_diff}, but for the results from night 1 with $K_\text{p} = 80$ km/s.}
    \label{fig_res_diff_2018_03_26_Kp_80}
\end{figure*}

\section{Results from injection and recovery tests for night 1 with \texorpdfstring{$\mathbf{K}_\text{p}~=~\mathbf{50}~$km/s}{Kp=50km/s}}\label{sec_res_night_1_Kp_50}

\begin{figure}
    \centering
    \includegraphics[width=\linewidth]{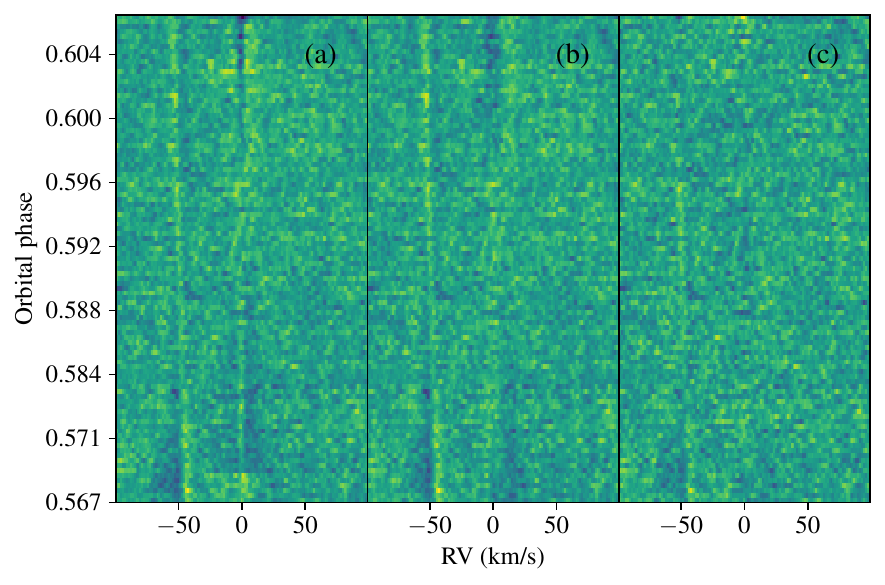}
    \caption{Same as Figure \ref{fig_planet_trail}, but for the results from night 1 with $K_\text{p} = 50$ km/s.}
    \label{fig_planet_trail_2018_03_26_Kp_50}
\end{figure}

\begin{figure*}
    \centering
    \includegraphics[width=\linewidth]{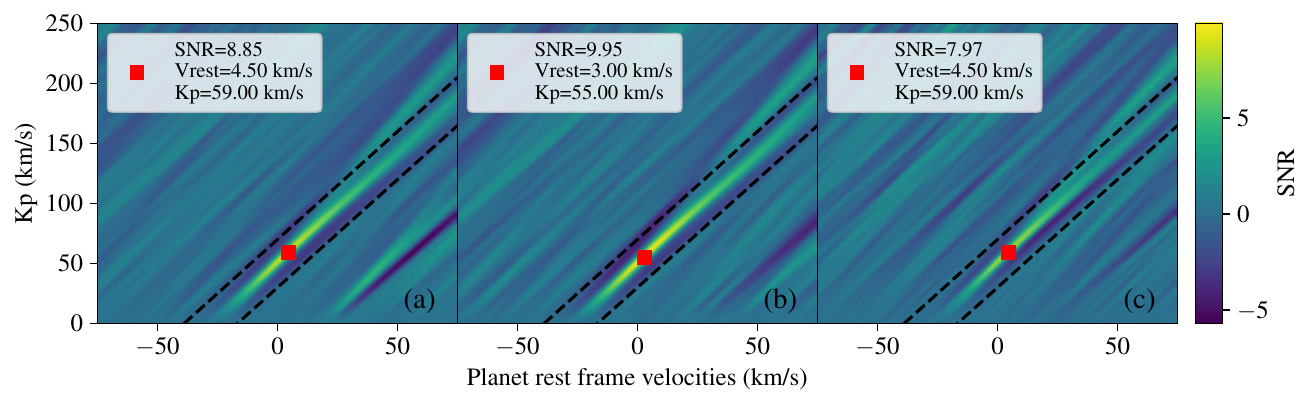}
    \caption{Same as Figure \ref{fig_velocity_map}, but for the results from night 1 with $K_\text{p} = 50$ km/s.}
    \label{fig_velocity_map_2018_03_26_Kp_50}
\end{figure*}

\begin{figure*}
    \centering
    \includegraphics[width=\linewidth]{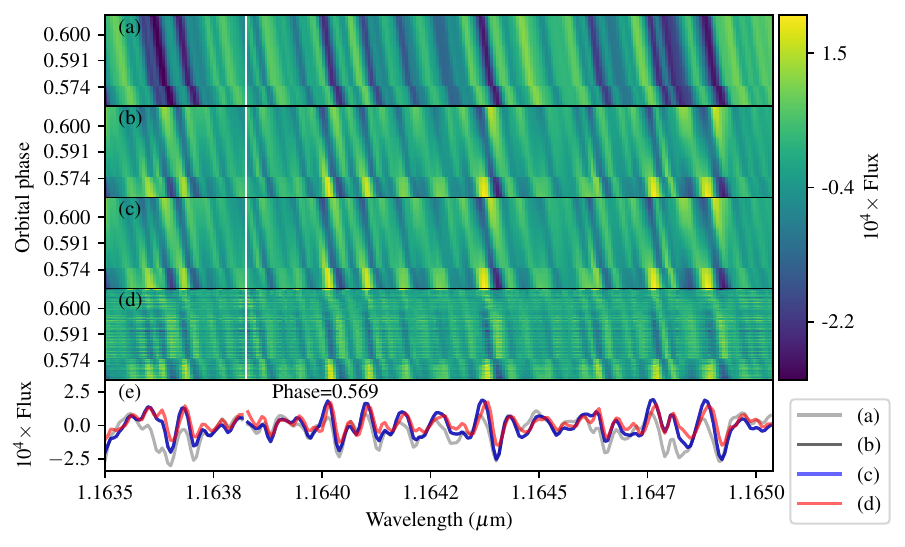}
    \caption{Same as Figure \ref{fig_res_diff}, but for the results from night 1 with $K_\text{p} = 50$ km/s.}
    \label{fig_res_diff_2018_03_26_Kp_50}
\end{figure*}

\section{Results from injection and recovery tests for night 2 with \texorpdfstring{$\mathbf{K}_\text{p}~=~\mathbf{110}~$km/s}{Kp=110km/s}}\label{sec_res_night_2_Kp_110}

\begin{figure}
    \centering
    \includegraphics[width=\linewidth]{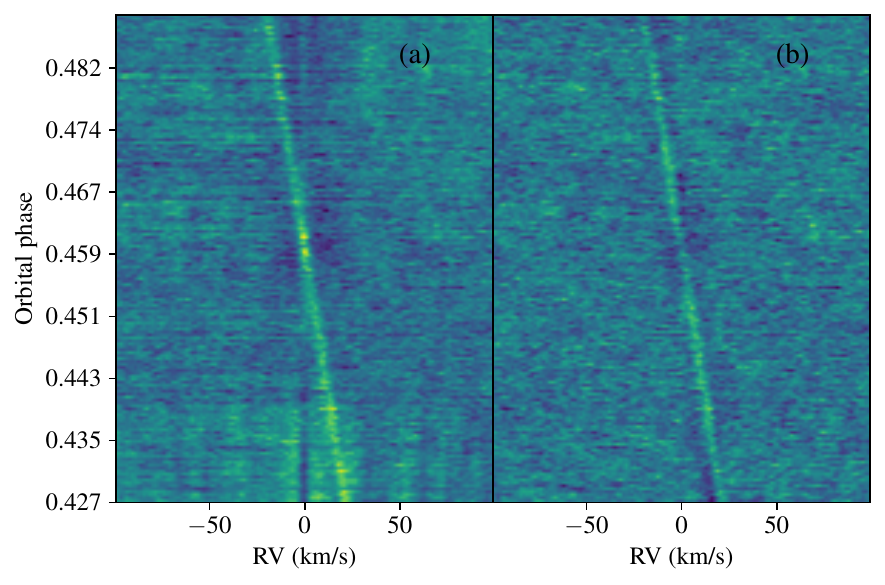}
    \caption{Same as Figure \ref{fig_planet_trail}, but for the results from night 2 with $K_\text{p} = 110$ km/s. Here, there are no Molecfit results, so panel (a) is from \textit{Astroclimes} and panel (b) is from PCA with $N_C~=~5$.}
    \label{fig_planet_trail_2019_03_15_Kp_110}
\end{figure}

\begin{figure*}
    \centering
    \includegraphics[width=\linewidth]{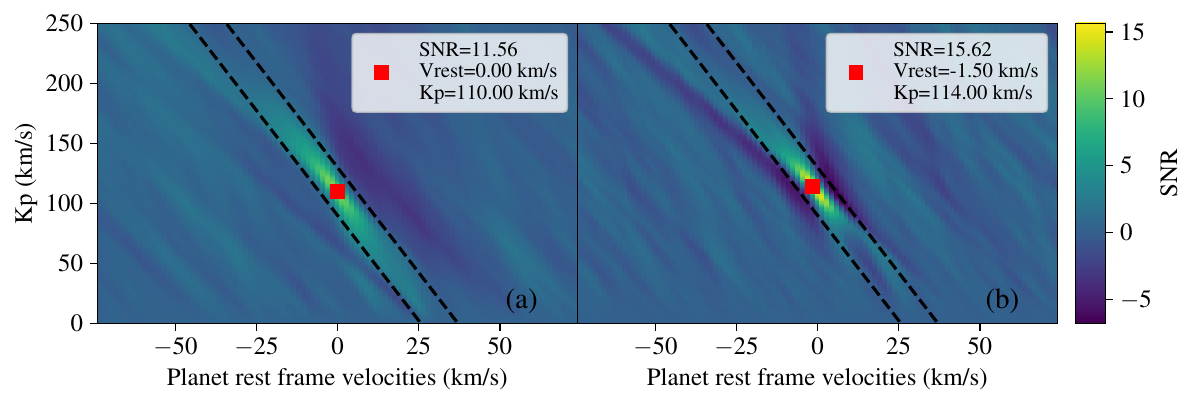}
    \caption{Same as Figure \ref{fig_velocity_map}, but for the results from night 2 with $K_\text{p} = 110$ km/s. Here, there are no Molecfit results, so panel (a) is from \textit{Astroclimes} and panel (b) is from PCA with $N_C~=~5$. The black dashed lines that define the ``signal'' and ``noise'' areas are now $y = -3.5V_\text{rest} + K_\text{p} \pm 20$, again determined by visual inspection.}
    \label{fig_velocity_map_2019_03_15_Kp_110}
\end{figure*}

\begin{figure*}
    \centering
    \includegraphics[width=\linewidth]{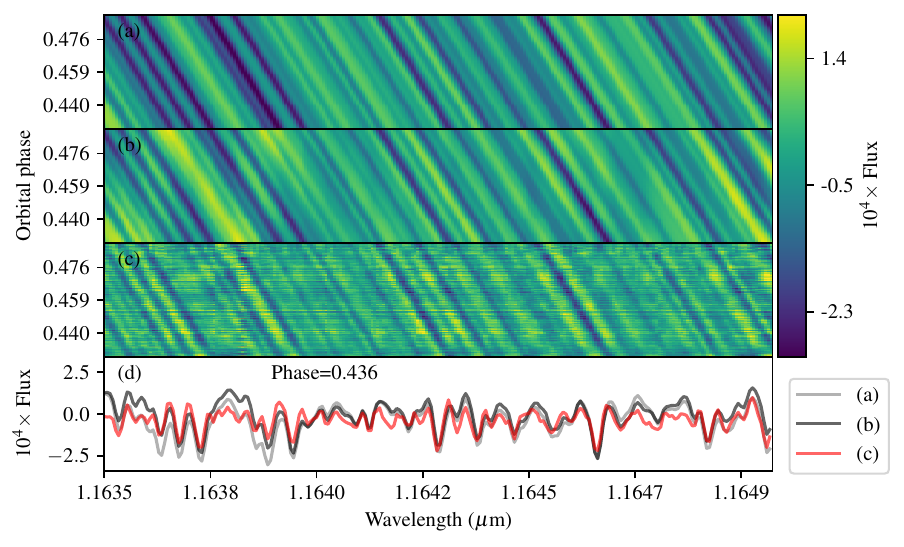}
    \caption{Same as Figure \ref{fig_res_diff}, but for the results from night 2 with $K_\text{p} = 110$ km/s. Here, there are no Molecfit results, so panel (b) is from \textit{Astroclimes} and panel (c) is from PCA with $N_C~=~5$.}
    \label{fig_res_diff_2019_03_15_Kp_110}
\end{figure*}



\bsp	
\label{lastpage}
\end{document}